\documentclass[twocolumn]{svjour3}

\usepackage{amsmath,amssymb}
\usepackage{graphicx}
\usepackage{epstopdf,textcomp}
\usepackage[usenames,dvipsnames]{xcolor}\usepackage{todonotes}

\journalname{Microfluidics and Nanofluidics}

\renewcommand\Im{\operatorname{Im}}

\begin{document}

\title{Energy conversion by surface-tension driven charge separation}

\author{Cesare Pini \and Tobias Baier \and Mathias Dietzel}
\institute{C. Pini, \email{cesare.pini@tu-dresden.de}, \at Polymeric Microsystems, TU Dresden, Helmholtzstr. 10, D-01069 Dresden, Germany \and 
					 T. Baier, \email{baier@csi.tu-darmstadt.de}, \and M. Dietzel, \email{dietzel@csi.tu-darmstadt.de}, 
					\at Center of Smart Interfaces, TU Darmstadt, Alarich-Weiss-Str. 10, D-64287 Darmstadt, Germany }
				
\date{Received: date / Accepted: date}

\maketitle


\begin{abstract}
In this work, the shear-induced electrokinetic streaming potential present in free-surface electrolytic flows subjected to a gradient in surface tension is assessed. Firstly, for a Couette flow with fully resolved electric double layer (EDL), the streaming potential per surface stress as a function of the Debye parameter and surface potential is analyzed. By contrast to the Smoluchowski limit in pressure-driven channel flow, the shear-induced streaming potential vanishes for increasing Debye parameter (infinitely thin EDL), unless the free surface contains (induced) surface charge or the flow at the charged, solid wall is permitted to slip. Secondly, a technical realization of surface-tension induced streaming is proposed, with surface stress acting on the free (slipping) surfaces of a micro-structured, superhydrophobic wall. The streaming potential is analyzed with respect to the slip parameter and surface charge. Finally, the surface tension is assumed to vary with temperature (thermocapillarity) or with surfactant concentration (destillocapillarity). The maximal thermal efficiency is derived and compared to the Carnot efficiency. For large thermal Marangoni number, the efficiency is severely limited by the large heat capacity of aqueous solvents. By contrast, destillocapillary flows may reach conversion efficiencies similar to pressure-driven flow.
\end{abstract}

\keywords{Electrokinetic energy conversion, \\Marangoni effect, Slip flow, Super-hydrophobic surface, Waste heat recovery, Thermodynamic analysis \\ 
\indent{\rule{7cm}{0.4pt}} \\ \\
This is the post-print authors' version of the manuscript, which was published in Microfluidics and Nanofluidics.
\copyright \ Springer-Verlag 2015. The final publication is available at Springer via 
http://dx.doi.org/10.1007/s10404-015-1597-x \\ \\
\indent{\rule{7cm}{0.4pt}}}



\section{Introduction}\label{Sec:intro}

The need for improving energy sustainability demands for the utilization of low-grade waste heat, which is emitted at temperatures just slightly above ambient. This can be attempted by various approaches, and many of them are in the focus of ongoing research. Examples are heat recuperation devices based on the Rankine cycle \cite{Chen:RenewSustainErgRev2010,Tchanche:RenewSustainErgRev2011}, adsorption refrigeration \cite{Saha:IJRefrig2003} or thermoelectric conversion \cite{Riffat:ApplThermEng2003,Bell:Science2008}. A major challenge is the circumstance that suitable techniques need to be available at low costs and low technical complexity: lower temperature levels imply smaller Carnot factors and, in order to harvest a non-negligible amount of exergy, massive parallelization of the waste-heat-recovery-devices has to be feasible. While efficient operation of most thermoelectric converters requires relatively high operating temperatures between $150$--$450 ^\circ\textrm{C}$ \cite{Saqr:ThermalScience2009} or higher \cite{Biswas:Nature2012}, the design and affordable fabrication of highly efficient thermoelectric converters is still topic of intense research \cite{Shakouri:AnnRevMatRes2011}. At this time, ultra-low-cost and robust low-complexity approaches of small-scale exergy recovery systems, which are applicable to any host device featuring a thermal gradient in a highly parallelizable fashion, are not readily available yet.

In this context, electrokinetic flow through micro-channels has received renewed attention as a means of converting kinetic energy of a flow driven by a pressure gradient into electric energy \cite{Yang:JMicroMechEng2003,vanderHeyden:PRL2005}. In such systems, free ions dissolved in a (typically electroneutral) carrier liquid accumulate in the vicinity of walls carrying a surface charge to form an electric double layer (EDL). Within this layer, typically a few up to a couple of hundred nm thick, the ions stay mobile and can be convectively transported along the channel as a streaming current. At steady-state, to comply to charge conservation, charge polarization induces a streaming potential, which in turn drives a conduction current equal in value but in opposite direction as the streaming current. Simple electrolytes come at low cost, and channel widths larger than the EDL-thickness are detrimental for the magnitude of the streaming potential. These characteristics render electrokinetic streaming suitable for miniaturization and parallelization so that sufficient power densities can be achieved. As a drawback, the performance of conventional electrokinetic streaming devices is spoiled by the circumstance that the excess ions accumulate in 
direct vicinity of fixed walls where frictional losses are highest. As a viable alternative one may consider flows where the excess ions screen externally applied electric fields at free surfaces not subject to the no-slip-condition. 

Work on electrokinetic free surface flows has mostly focused on the interaction of external electric fields with the fluid domain, either to address stability issues \cite{Taylor:JFM1965,Neron:JPhysIIFrance} and electro-osmotic propulsion of liquid films \cite{Melcher:AnnRevFluidMech1969} or to understand electrosprays \cite{Salata:CurrNanoscie2005} as well as electrowetting \cite{Mugele:JPhysCondMatt2005}, to name a few. In recent years, along with the general trend of miniaturization, flow domains of similar characteristic length scale as the EDL-thickness have received increased interest \cite{Qian:MechResComm2009}. Such considerations are crucial in drainage models of thin films and foams \cite{Tsekov:Langmuir2010,Karakashev:Langmuir2011} or to address the electrohydrodynamic stability of ultrathin electrolyte films which are electro-osmotically sheared by an external field \cite{Joo:JMechScieTech2008,Mayur:PhysRevE2012}. In comparison, much less work has been done on mechanically driven charge separation in flow domains where at least one interface is not bounded by a wall. The most prominent example of this category is probably Lord Kelvin's famous water dropper to generate direct current (DC) voltages in the kV-range \cite{Thomson:ProcRSocL1867,Marin:LabChip2013}. The high voltages are achieved by the circumstance that the convectively transported charges are enclosed within drops, which are (electrically) insulated from each other by (dielectric) air. While the continuous domain in conventional electrokinetic streaming in channels limits the streaming potential by the opposing conduction current, the latter is avoided altogether in the Kelvin dropper and related devices \cite{Duffin:JPhysChemC2008}. With respect to continuous electrokinetic free surface flows driven by a pressure gradient, researchers have addressed charge separation in free-surface guided microchannels \cite{Lee:JAP2006} as well as electrokinetic flow over superhydrophobic surfaces \cite{Zhao:PoF2011,Seshadri:PoF2013}. In the latter, wall friction is reduced by suspending the flow on an array of air pockets trapped in the channel walls. 

Free surface flows can be driven also by stresses at the interface caused by a non-uniform surface tension. Corresponding effects become particularly dominant for large surface-to-volume ratios of the fluidic domain. Surface tension is affected by temperature (thermocapillarity or thermal Marangoni effect) or by the concentration of another dissolved phase (destillocapillarity or solutal Marangoni effect). The shear-induced electrokinetic streaming in free surface flows with the aim to (partially) convert thermal or chemical energy into electric energy is in the focus of the present study.  

In section \ref{Sec:Couette}, double layer effects are addressed at hand of a (hypothetical) Couette-type of flow, including those caused by a molecular wall slip of similar order as the Debye-length. In section \ref{Sec:channel}, as a technically more feasible example and in the limit of infinitely thin EDL, previous work on electrokinetic flow over superhydrophobic surfaces is extended to account for fluid propulsion by means of a surface tension gradient along the grooves, which enclose air pockets. Based on these considerations, in section \ref{Sec:thermocap}, the thermal efficiency is derived when the surface tension is a function of temperature (thermal Marangoni effect), i.e. thermocapillarity provides for the required liquid propulsion. The latter system is one of the first, technically feasible approaches of thermally driven electrokinetic charge separation \cite{Grosu:SurfEngApplElectrochem2010}, directly converting thermal into electric energy. It is a low-cost and low-complexity approach and might be useful as small-scale waste heat recovery device.

\section{Electrokinetic streaming in planar Couette-type flow}\label{Sec:Couette}

\begin{figure}
	\centerline{\includegraphics[width=8.4cm]{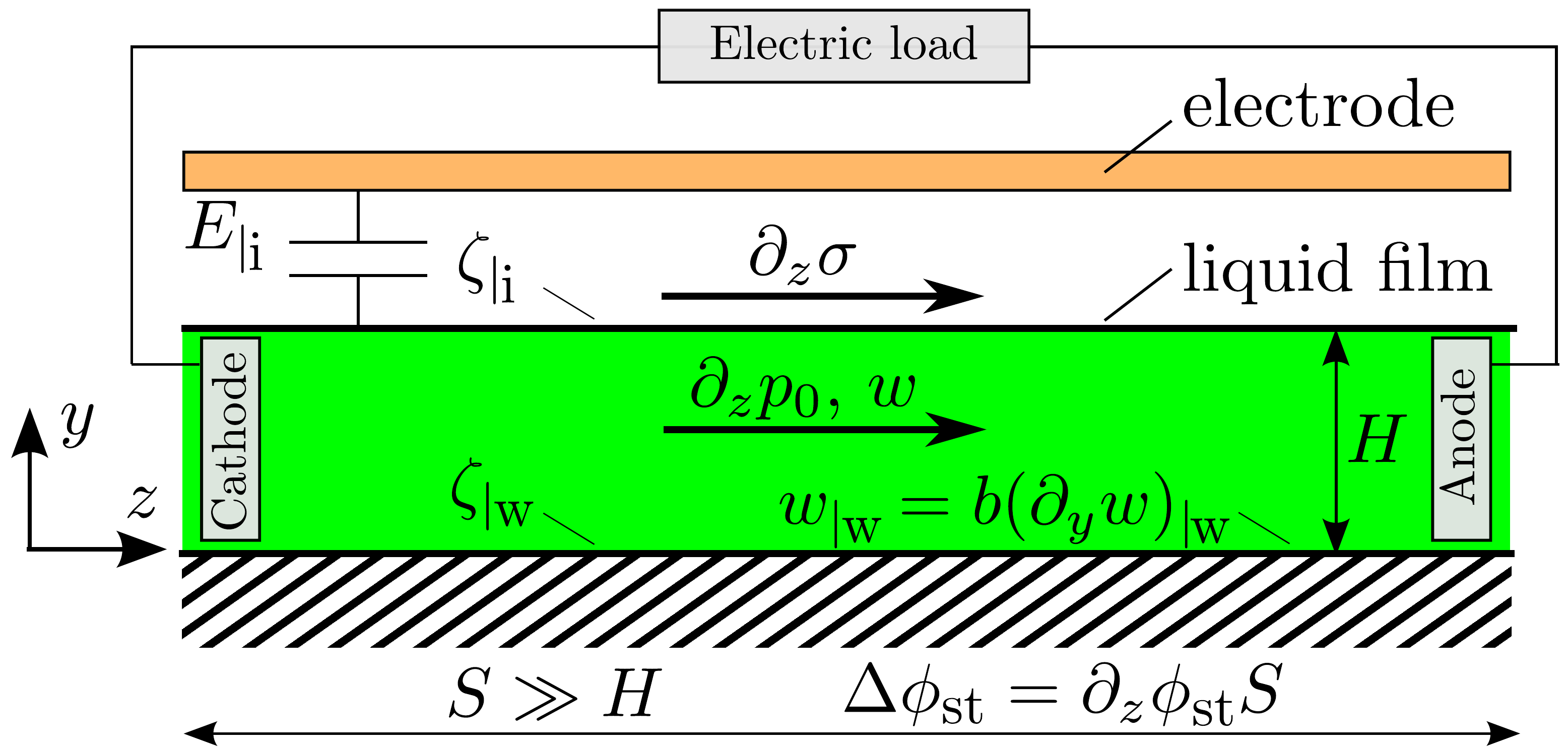}}
	\caption{Sketch of the planar liquid film of film height $H$, horizontal extent $S \gg H$ and driven by a gradient in surface tension, $\partial_z \sigma$, as well as by a pressure 
	gradient $\partial_z p_0$. The propulsion leads to an axial velocity distribution $w(y)$ and a streaming potential of $\Delta \phi_{\textrm{st}} = \partial_z \phi_{\textrm{st}} S$. The wall has a $\zeta$-potential of $\zeta_{|\textrm{w}}$, 	while the potential at the interface is denoted by $\zeta_{|\textrm{i}}$. The potentials are related to electric fields $E_{|\textrm{w}}$ (or surface charge density $q_{|\textrm{w}}$) and $E_{|\textrm{i}}$ (or $q_{|\textrm{i}}$), respectively.}
	\label{fig:setup_couette}
\end{figure}

As schematically shown in figure \ref{fig:setup_couette}, in this section a liquid layer of a symmetric electrolyte of thickness $H$ (in $y$-direction) is considered. The horizontal extent $S$ in spread direction $z$ (from now on termed the axial direction) of the layer is assumed to be much larger than $H$. The flow is viewed as being uniform in $x$-direction, i.e. the system is essentially two-dimensional. The layer is bounded from below by a flat solid wall (subscript '$w$'), exhibiting a Navier-slip coefficient $b$, and from above by an inert gas phase. The liquid-gas interface (subscript '$i$') has a surface tension $\sigma$. The latter varies in $z$-direction, giving raise to a shear-induced fluid propulsion and streaming potential $\phi_{\textrm{st}}(z)$. The flow is potentially supported by an axial pressure gradient, $\partial_z p$, where $\partial_z (.) = \partial (.)/\partial z$. In the following, the electric double layer (EDL), with a thickness of $\lambda_{\textrm{D}}$ typically in the order of $1-10\:\textrm{nm}$, will be resolved. This is only useful if $H$ is at least not much larger than $\lambda_{\textrm{D}}$. Such ultra-thin films are known to dewett most solid substrates or at least undergo significant surface deformations. Here it is assumed that the solid surface is treated in such a fashion that it is superhydrophilic to polar liquids \cite{Drelich:SoftMatter2011} and remains fully wetted at all times. For instance, for an aqueous solution this can be accomplished by coating a surface with titanium dioxide ($\textrm{Ti}\textrm{O}_2$) and subsequent irradiation with ultra-violet (UV) light \cite{Watanabe:ThinSolidFilms1999}. Furthermore, as will be discussed in the next section, an electric field perpendicular to the film interface will be applied to induce an interfacial charge. The corresponding electrohydrodynamic pressure generated at the interface is negative and lowers the local fluid pressure. This leads to thickening of the film, i.e. it counteracts dewetting effects, at least if the applied electric field is below the threshold of electrohydrodynamics instabilities at the interface \cite{Taylor:JFM1965}. Thus, the surface deformations will be neglected and the layer has a uniform and constant thickness $H$.

\subsection{Axial velocity distribution}\label{Subsec:velprof}
The liquid motion is described by the velocity vector $\vec{v} = (0,v,w)$, where incompressibility $\nabla \cdot \vec{v} = 0$ holds. Furthermore, assuming Newtonian, low Reynolds number (creeping) flow of (constant) viscosity $\eta$, the Navier-Stokes equation in $z$-direction can be approximated by (Stokes limit)
\begin{equation}
\label{Eq:NSEax} 0 = -\partial_z p + \eta \nabla^2 w - \rho_e \partial_z \phi,
\end{equation}
where $p$ denotes the total fluid pressure. Gravitational effects are neglected. The flow is fully developed so that $\nabla^2 w = \partial^2_y w$. The electrostatic forces on the (neutral) solvent due to the dissolved ions are considered by the Maxwell stresses in terms of the charge density $\rho_e$ and the total electric potential $\phi$. The latter is the linear superposition of an electric double layer potential $\psi(y,z)$ and the streaming potential, i.e. $\phi(y,z) = \psi(y,z) + \phi_{\textrm{st}}(z)$. The charge density can be expressed by the Poisson equation. Using the lubrication approximation $(H/S)^2 \ll 1$, $\rho_e$ is given by
\begin{equation}
\label{Eq:Poisson} \rho_e \approx -\epsilon \partial^2_y \psi,
\end{equation}
where $\epsilon$ is the (constant) dielectric permittivity. With this, the momentum balance in $y$-direction can be expressed by
\begin{equation}
\label{Eq:NSEvert} p = p_0 + \epsilon \frac{1}{2} (\partial_y \psi)^2,
\end{equation}
where $p_0(z)$ is the externally applied hydraulic pressure. The last term on the right-hand side (RHS) is the electrohydrostatic pressure. In what follows, the variation of $\psi$ in streaming direction $z$ is omitted. Inserting (\ref{Eq:Poisson}) and (\ref{Eq:NSEvert}) into (\ref{Eq:NSEax}) and integrating twice in $y$-direction leads to the velocity distribution $w(y,z)$ in (axial) streaming direction, namely
\begin{align} \label{Eq:velax}
w =& \frac{\partial_z p_0}{\eta}\left[\frac{1}{2} y^2 - H (y+b)\right] + \frac{\partial_z \sigma}{\eta}(y+b) \nonumber \\
&- \frac{\epsilon \partial_z \phi_{\textrm{st}}}{\eta}\left[\psi - \zeta_{|\textrm{w}} + y E_{|\textrm{i}} + b(E_{|\textrm{i}}-E_{|\textrm{w}})\right].
\end{align}
In (\ref{Eq:NSEax}), for shorter notation, the electric fields $E_{|\textrm{w}} \equiv - (\partial_y \psi)_{|\textrm{w}}$ and $E_{|\textrm{i}} \equiv - (\partial_y \psi)_{|\textrm{i}}$ were used as well as $\psi_{|\textrm{w}} \equiv \zeta_{|\textrm{w}}$ (charges in the Stern-layer are ignored). The subscript $|w$ stands for evaluation at $y=0$. In (\ref{Eq:velax}), the Navier-slip condition $w_{|\textrm{w}} = b (\partial_y w)_{|\textrm{w}}$ applies at the wall. At the sheared interface, the stress condition $\eta(\partial_y w)_{|\textrm{i}} = \partial_z \sigma$ is fulfilled, where the subscript $|i$ stands for evaluation at $y=H$.

As it becomes relevant in section \ref{Sec:thermocap}, (\ref{Eq:velax}) remains valid if the viscosity is dependent on temperature and the latter varies in $z$ only. In this case, as a consequence of the expansion in terms of $H/S$ within a lubrication approximation, contributions to (\ref{Eq:NSEax}) due to $\nabla \eta$ emerging from the complete viscous stress term, $\nabla \cdot \eta [\nabla \vec{v} + (\nabla \vec{v})^T]$, can be shown to be of higher order in $H/S$ and negligible \cite{Dietzel:JFM2015}.

\subsection{Electric currents}\label{Subsec:currents}

Using the Navier-slip condition, the (width-averaged) streaming current $I_{\textrm{st}}\:=\:W^{-1}\int^W_0\int^{H}_0 \rho_e\:w\:dy\:dx$ can be expressed by
\begin{align} \label{Eq:streamcurr}
&\frac{I_{\textrm{st}}}{\epsilon} = \nonumber \\
&-\frac{\partial_z p_0}{\eta}\Bigg[\int^{H}_0 \psi dy\!-\! H \zeta_{|\textrm{w}}\! +\! \frac{1}{2} H^2 E_{|\textrm{i}}\! +\! b H (E_{|\textrm{i}}-E_{|\textrm{w}})\Bigg] \nonumber \\
&+ \frac{\partial_z \sigma}{\eta}\left[\zeta_{|\textrm{i}}\!-\! \zeta_{|\textrm{w}}\!+\! H E_{|\textrm{i}}\!+\! b(E_{|\textrm{i}}\!-\! E_{|\textrm{w}}) \right] \nonumber \\
&- \frac{\epsilon \partial_z \phi_{\textrm{st}}}{\eta}\Bigg[\int^{H}_0 (\partial_y \psi)^2 dy\!+\! 2 E_{|\textrm{i}}(\zeta_{|\textrm{i}}\!-\! \zeta_{|\textrm{w}})\!+\! H E^2_{|\textrm{i}} \nonumber \\
&+ b (E_{|\textrm{i}}\!-\! E_{|\textrm{w}})^2 \Bigg],
\end{align}
where $W$ is the extent of the layer in $x$-direction.

Along with the condition of zero ion flux across the interface, using the Nernst-Planck equation in the lubrication approximation suggests that each ion species obeys a Boltzmann distribution, namely
\begin{equation}
\label{Eq:Boltzmann} n_k = n_{k|\textrm{c}} \textrm{exp}\left[-\overline{\nu}_k \left(\psi - \psi_{|\textrm{c}} \right)/\psi_{\textrm{D}}\right],
\end{equation}
where $\psi$ is the EDL-potential and $n_k$ ($k=1...K$) are the local ion number concentrations, while $n_{k|\textrm{c}}$ and $\psi_{|\textrm{c}}$ are the concentrations at an (arbitrary) reference location $y=y_{|\textrm{c}}$ further away from charged walls and interfaces. Furthermore, one has $\psi_{\textrm{D}} = k_{\textrm{B}} T_\infty/(e \nu)$ and $\overline{\nu}_k = \nu_k/\nu$, where the elementary charge is denoted by $e$, $\nu_k$ are the ion valences, $k_{\textrm{B}}$ is the Boltzmann constant and $T$ the temperature. A non-uniform temperature either compresses or expands the EDL \cite{Dietzel:muFlu2012}. For $T=T(z)$, $n_{k|\textrm{c}}$ may vary in $z$, giving rise to variations of $\psi$ (and $\psi_{|\textrm{c}}$) in $z$ as well. Besides affecting the ion distribution and the conduction current, this can have an effect on the mechanical equilibrium of the ion cloud (thermo-osmosis) \cite{Dietzel:JFM2015}, which is beyond the scope of the present study. Corresponding effects were found to be generally small and only noticeable if the liquid film thickness is of the same order as the EDL. Thus, in the present work we neglect the (weak) dependence of $\psi$ on $T$, which was already implied by using $\partial_z \psi = 0$ before. Since $\rho_e = e \sum^K_{k=1} \nu_k n_k$, (\ref{Eq:Poisson}) suggests that in this case also $\partial_z n_k = 0$. It follows that $n_{k|\textrm{c}} \equiv n_\infty$, $\psi_{|\textrm{c}} \equiv 0$ and $T \equiv T_\infty$, where $n_\infty$ is a constant reference concentration and $T_\infty$ is a constant reference temperature.

In this case, the only non-convective mechanism of ion transport in streaming direction is the total conduction current due to the induced potential gradient, reading
\begin{equation}
\label{Eq:condcurr} I_{\textrm{cd}} = -e^2 \partial_z \phi_{\textrm{st}} \int^{H}_0 \sum^K_{k=1} \omega_{n,k} \nu^2_k n_k dy,
\end{equation}
where $\omega_{n,k} \approx D_{\textrm{n},k}/(k_{\textrm{B}} T)$ denote the ionic mobilities, with $D_{\textrm{n},k}$ as the (Fickian) diffusion coefficients. With this, the conduction current of a symmetric $\nu:\nu$ electrolyte with identical diffusion coefficient $D_{\textrm{n},k} \equiv D_{\textrm{n}}$ for each ion species can be written as
\begin{equation}\label{Eq:condcurr2} 
\frac{I_{\textrm{cd}}}{\epsilon} = -\frac{D_{\textrm{n}} \partial_z \phi_{\textrm{st}}}{\lambda^2_{\textrm{D}}} \frac{T_\infty}{T} F_{\textrm{CS}}
\end{equation}
where 
\begin{equation}\label{Eq:condfct} 
F_{\textrm{CS}} = \int^{H}_0 \textrm{cosh}\left(\psi/\psi_{\textrm{D}} \right) dy
\end{equation}
and
\begin{equation}\label{Eq:EDLlength}
\lambda_{\textrm{D}} = \sqrt{\frac{\epsilon k_{\textrm{B}} T_\infty}{2 e^2 \nu^2 n_\infty}} 
\end{equation}
is the (nominal) EDL-thickness.

\subsection{Streaming potential}\label{Subsec:streaming}

\subsubsection{General considerations}\label{Subsubsec:general}

By requiring that the total current $I_{\textrm{st}} + I_{\textrm{cd}}$ vanishes, charge conservation determines the convection-induced streaming potential, leading to
\begin{align}
\label{Eq:streampot}
&-\partial_z \phi_{\textrm{st}} = \nonumber \\
&\Bigg\{-\partial_z p_0 \bigg[H \zeta_{|\textrm{w}}\!-\!\int^{H}_0 \psi dy\!-\!\frac{1}{2} H^2 E_{|\textrm{i}}\!+\! b H (E_{|\textrm{w}}-E_{|\textrm{i}})\bigg] \nonumber \\
&+ \partial_z \sigma \left[\zeta_{|\textrm{w}}\,-\,\zeta_{|\textrm{i}}\!-\! H E_{|\textrm{i}}\!+\! b(E_{|\textrm{w}}\!-\! E_{|\textrm{i}}) \right]\Bigg\}/F_{\phi},
\end{align}
where
\begin{equation}
\label{Eq:denominator}
\begin{aligned}
F_{\phi} &= \frac{\eta D_{\textrm{n}}}{\lambda^2_{\textrm{D}}} \frac{T_\infty}{T} F_{\textrm{CS}} + \epsilon\Bigg[\int^{H}_0 (\partial_y \psi)^2 dy - 2 E_{|\textrm{i}}(\zeta_{|\textrm{w}}-\zeta_{|\textrm{i}}) \\
&+ H E^2_{|\textrm{i}} + b (E_{|\textrm{w}}-E_{|\textrm{i}})^2 \Bigg].
\end{aligned}
\end{equation}
Expression (\ref{Eq:streampot}) remains valid if $T$ varies in $z$-direction. In this case, in (\ref{Eq:denominator}) the local viscosity and temperature needs to be used. However, the hydrodynamic radius $R_0$ of common salt ions is relatively unaffected by temperature \cite{Oelkers:JSolChem1989}. Then, the Stokes-Einstein-relation $D_{\textrm{n}} \approx k_{\textrm{B}} T /(6 \pi \eta R_0)$ implies that $\eta D_{\textrm{n}} T_\infty/T = \eta_\infty D_{\infty,\textrm{n}}$ is a constant, where $D_{\infty,\textrm{n}}$ and $\eta_\infty$ are the diffusion coefficient and the liquid viscosity, respectively, both determined at $T = T_\infty$. Hence, at least for simple 1:1-electrolytes, (\ref{Eq:denominator}) and (\ref{Eq:streampot}) can be considered to be unaffected in case that $T=T(z)$. This will become relevant in section \ref{Sec:thermocap}.

For evaluation of the streaming potential described by (\ref{Eq:streampot}), an expression for the EDL-potential need to be found which fulfills the Poisson-Boltzmann equation. With (\ref{Eq:Boltzmann}), the latter reads
\begin{equation}
\label{Eq:PBeq}
\partial^2_y \left(\psi/\psi_{\textrm{D}}\right) = \lambda^{-2}_{\textrm{D}} \textrm{sinh}\left(\psi/\psi_{\textrm{D}}\right).
\end{equation}
In the vicinity of a boundary with arbitrary values of the potential $\zeta_{|\textrm{s}}$, the Gouy-Chapman model (GC) allows for an analytical solution of (\ref{Eq:PBeq}), while the overlap of the EDLs is neglected. As it will be shown, this inaccuracy is only relevant if $H$ is equal or smaller than $\lambda_{\textrm{D}}$ and $\psi/\psi_{\textrm{D}}$ exceeds ${\cal O} (1)$ at the same time. The GC-model is given by \cite{Russel:Cambridge1989} (page 102)
\begin{equation}
\label{Eq:EDLpotChapman}
\frac{\psi}{\psi_{\textrm{D}}} = 2 \textrm{ln}\left[\frac{1+\textrm{exp}(-\widetilde{y}/\lambda_{\textrm{D}})\textrm{tanh}
(\frac{1}{4}\zeta_{|\textrm{s}}/\psi_{\textrm{D}})}{1-\textrm{exp}(-\widetilde{y}/\lambda_{\textrm{D}})\textrm{tanh}(\frac{1}{4}\zeta_{|\textrm{s}}/\psi_{\textrm{D}})}\right], 
\end{equation}
where $\widetilde{y} \geq 0$ is directed normal from the charged boundary into the interior of the electrolyte. For the potential due to the wall charge, $\widetilde{y} \equiv y$ and the subscript $|s \equiv |w$, while for the interfacial charge, $\widetilde{y} \equiv H - y$ and $|s \equiv |i$. The $\zeta$-potentials and electric fields are not independent from each other but related via the surface charge densities $q = -\epsilon \partial_y \psi (\vec{j} \cdot \vec{n}_{\textrm{B}})$, where $\vec{j}$ is the unit vector in $y$-direction and $\vec{n}_{\textrm{B}}$ is the outward directed normal vector of the domain boundary. Accordingly, at the wall one has $q_{|\textrm{w}} = \epsilon E_{|\textrm{w}} (\vec{j} \cdot \vec{n}_{\textrm{B}})$ and at the interface $q_{|\textrm{i}} = \epsilon E_{|\textrm{i}} (\vec{j} \cdot \vec{n}_{\textrm{B}})$. The derivative of (\ref{Eq:EDLpotChapman}) leads to an expression for $\zeta_{|\textrm{s}}$ as a function of the corresponding surface charge density $q_{|\textrm{s}}$, namely
\begin{equation}
\label{Eq:zeta_surfcharge}
\zeta_{|\textrm{s}} = -2 \psi_{\textrm{D}} \textrm{arsinh}\left(\frac{\lambda_{\textrm{D}}}{2 \epsilon \psi_{\textrm{D}}} q_{|\textrm{s}}\right).
\end{equation}
Non-overlapping EDLs imply that far away from the charged boundary, $(\partial_y \psi)_{|\infty} \rightarrow 0$ and $(\psi)_{|\infty} \rightarrow 0$. Along with (\ref{Eq:PBeq}), this can be used to show that
\begin{equation}
\label{Eq:condfct2} 
F_{\textrm{CS}} = H + \frac{\lambda^2_{\textrm{D}}}{2} \int^{H}_0 \left(\frac{\partial_y \psi}{\psi_{\textrm{D}}}\right)^2 dy.
\end{equation}
Subsequently, using (\ref{Eq:EDLpotChapman}), one has
\begin{align}
\label{Eq:EDLfieldChapman}
&\int^{H}_0 \left(\frac{\partial_y \psi}{\psi_{\textrm{D}}}\right)^2 dy = \frac{8}{\lambda_{\textrm{D}}} \left[\textrm{exp}(-2 H/\lambda_{\textrm{D}}) - 1\right] \times \nonumber \\
&\mkern63mu\Bigg[\frac{\textrm{sinh}^2(\frac{1}{4} \zeta_{|\textrm{i}}/\psi_{\textrm{D}})}{\textrm{exp}(-2 H/\lambda_{\textrm{D}})\textrm{tanh}^2(\frac{1}{4} \zeta_{|\textrm{i}}/\psi_{\textrm{D}})-1} \nonumber \\
&\mkern63mu+ \frac{\textrm{sinh}^2(\frac{1}{4} \zeta_{|\textrm{w}}/\psi_{\textrm{D}})}{\textrm{exp}(-2 H/\lambda_{\textrm{D}})\textrm{tanh}^2(\frac{1}{4} \zeta_{|\textrm{w}}/\psi_{\textrm{D}})-1}\Bigg].
\end{align}
The corresponding $\zeta$-potentials as a function of the surface charge densities can be evaluated with (\ref{Eq:zeta_surfcharge}). Hence, $F_{\phi}$ expressed by (\ref{Eq:denominator}) is fully determined. Finally, integration of (\ref{Eq:EDLpotChapman}) leads to
\begin{align}
\label{Eq:integratedEDLpot}
\int^{H}_0 \frac{\psi}{\psi_{\textrm{D}}} dy = \lambda_{\textrm{D}} \Bigg\{
&\chi_{|\textrm{w}} \Lambda(\chi^2_{|\textrm{w}},2,\frac{1}{2}) 
+\chi_{|\textrm{i}} \Lambda(\chi^2_{|\textrm{i}},2,\frac{1}{2}) \nonumber\\
-&\hat{\chi}_{|\textrm{w}} \Lambda(\hat{\chi}^2_{|\textrm{w}},2,\frac{1}{2}) 
-\hat{\chi}_{|\textrm{i}} \Lambda(\hat{\chi}^2_{|\textrm{i}},2,\frac{1}{2})
\Bigg\},
\end{align}
where $\chi_{|\textrm{w}/\textrm{i}} = \textrm{tanh}(\frac{1}{4} \zeta_{|\textrm{w}/\textrm{i}}/\psi_{\textrm{D}})$, $\hat{\chi}_{|\textrm{w}/\textrm{i}}=\chi_{|w/i}e^{-H/\lambda_{\textrm{D}}}$ and
\begin{equation}
\label{Eq:Lechfct}
\Lambda(\varrho,\vartheta,\xi) = \sum^\infty_{k=0} (\xi+k)^{-\vartheta} \varrho^k
\end{equation}
is the Lerch transcendent. In the derivation of (\ref{Eq:integratedEDLpot}), the identity \cite{Gradshteyn:2007}
\begin{equation}
\label{Eq:intformula_GR_2.728_2}
\int \textrm{ln}\left( 1 + a z\right) \frac{dz}{z} = (az) \Lambda( -az, 2, 1)
\end{equation}
and $\Lambda(\varrho^2,2,1/2) = 2[\Lambda(\varrho,2,1) + \Lambda(-\varrho,2,1)]$ was used. With (\ref{Eq:EDLfieldChapman}) and (\ref{Eq:integratedEDLpot}), the streaming potential (\ref{Eq:streampot}) as a function of the viscosity, diffusivity, bulk salt concentration, surface charge densities as well as EDL- and film thicknesses is fully determined.

At small $\zeta$-potentials but arbitrary film heights, the results obtained with the GC-model can be compared to those found with the Debye-H\"uckel (DH) approximation. In general, one may specify the EDL-potential in terms of either the surface charge densities or the $\zeta$-potentials at the boundaries. Although the first option might be in many cases physically more meaningful, the GC-model is derived by specifying the relevant $\zeta$-potential. To allow for easier comparison between the models used in this work and also with 
other work of electrokinetic streaming found in the literature, we will follow this approach in the DH-model as well. To this end, two cases will be distinguished, one where only one wall $\zeta$-potential is applied (while $(\partial_y \psi)_{|\infty} \rightarrow 0$ and $(\psi)_{|\infty} \rightarrow 0$), and another, where $\zeta_{|\textrm{w}}$ as well as $\zeta_{|\textrm{i}}$ are imposed. The first is simply described by
\begin{equation}
\label{Eq:EDLpot_DH_eqGC}
\psi^{(\textrm{DH,NO})} = \zeta_{|\textrm{w}} \textrm{exp}(-y/\lambda_{\textrm{D}})
\end{equation}
and is equivalent to the GC-model of non-overlapping (NO) EDLs. This implies that for very thin films, the charges inside the EDL might not completely screen the wall surface charge, i.e. a finite electric field at the interface might remain. The second case is governed by
\begin{align}
\label{Eq:EDLpot_DH}
\psi^{(\textrm{DH})} = \zeta_{|\textrm{i}} \frac{\textrm{sinh}(y/\lambda_{\textrm{D}})}{\textrm{sinh}(H/\lambda_{\textrm{D}})} + \zeta_{|\textrm{w}} \frac{\textrm{sinh}[(H-y)/\lambda_{\textrm{D}}]}{\textrm{sinh}(H/\lambda_{\textrm{D}})}.
\end{align}
Here, the electric fields at the boundaries are derived from (\ref{Eq:EDLpot_DH}) and given by
\begin{equation}
\label{Eq:field_i_DH}
E^{(\textrm{DH})}_{|\textrm{i}} = -\frac{\zeta_{|\textrm{i}}}{\textrm{tanh}(H/\lambda_{\textrm{D}})\lambda_{\textrm{D}}} + \frac{\zeta_{|\textrm{w}}}{\textrm{sinh}(H/\lambda_{\textrm{D}}) \lambda_{\textrm{D}}}
\end{equation}
and
\begin{equation}
\label{Eq:field_w_DH}
E^{(\textrm{DH})}_{|\textrm{w}} = -\frac{\zeta_{|\textrm{i}}}{\textrm{sinh}(H/\lambda_{\textrm{D}})\lambda_{\textrm{D}}} + \frac{\zeta_{|\textrm{w}}}{\textrm{tanh}(H/\lambda_{\textrm{D}}) \lambda_{\textrm{D}}},
\end{equation}
respectively. Just as for the Gouy-Chapman case, the corresponding streaming potential is inferred from (\ref{Eq:streampot}) for both cases.

As will be used in section \ref{Sec:channel}, for large $H \gg \lambda_{\textrm{D}}$ at arbitrary slip length $b$, the streaming potential is found to be
\begin{align}
\label{Eq:streampot_largeh0_general}
&-(\partial_z \phi_{\textrm{st}})_{H \gg \lambda_{\textrm{D}}} = \nonumber \\
&\frac{-\partial_z p_0 H \left[\frac{1}{2} q_{|\textrm{i}}\!+\! \frac{b}{H}(q_{|\textrm{i}}\!+\! q_{|\textrm{w}})\right]\!+\! \partial_z\sigma \left[q_{|\textrm{i}}\!+\! \frac{b}{H}(q_{|\textrm{i}}\!+\! q_{|\textrm{w}})\right]}
{\eta_\infty \sigma^\infty\!+\! q^2_{|\textrm{i}}\!+\! \frac{b}{H}(q_{|\textrm{i}}+q_{|\textrm{w})^2}}
\end{align}
where $\sigma^\infty = 2 e^2 \nu^2 n_\infty D_{\infty,\textrm{n}}/(k_{\textrm{B}} T_\infty) = \epsilon D_{\infty,\textrm{n}}/\lambda^2_{\textrm{D}}$ is the electric conductivity of the bulk electrolyte. In (\ref{Eq:streampot_largeh0_general}), since by assumption $H \gg \lambda_{\textrm{D}}$, the exact form of the ion distribution in the double layer is irrelevant. Hence, within this limit of infinitely thin EDLs and no wall slip, a shear-induced streaming potential (with $\partial_z p_0 = 0$) is solely caused by the convective motion of charges accumulated at the free surface not bounded by the no-slip condition.

\subsubsection{Streaming potential without induced interfacial charge and wall slip}\label{Subsubsec:no_ifcharge_wallslip}

In the following, the streaming potential is analyzed for the case that only a fixed wall $\zeta$-potential equal to $\zeta_{|\textrm{w}}$ is present. Furthermore, the fluid at the wall is assumed to comply to the no-slip-condition. Then, the streaming potential reads
\begin{equation}
\label{Eq:streampot_noslip_nofield}
-\partial_z \phi_{\textrm{st}} = \frac{-\partial_z p_0 \left(H \zeta_{|\textrm{w}}\!-\! \psi_{\textrm{D}} \int^{H}_0\! \frac{\psi}{\psi_{\textrm{D}}}\! dy \right)
\!+\! \partial_z \sigma \zeta_{|\textrm{w}}}{\frac{\eta_\infty D_{\infty,\textrm{n}}}{\lambda^2_{\textrm{D}}}H\!+\! \left(\frac{\eta_\infty D_{\infty,\textrm{n}}}{2}
\!+\! \epsilon \psi^2_{\textrm{D}}\right) \int^{H}_0\!\left(\frac{\partial_y \psi}{\psi_{\textrm{D}}}\right)^2\!dy}.
\end{equation}
From this one can see that there is a qualitative difference between shear- and pressure-induced electrokinetic streaming. For large $H$, the integrals in (\ref{Eq:streampot_noslip_nofield}) vanish so that, on the one hand, the pressure-induced streaming potential (with $\partial_z \sigma$ set to zero) attains the well-known Smoluchowski limit given by 
\begin{equation}
\label{Eq:streampot_Smoluchowski}
\left(\frac{\partial_z \phi_{\textrm{st}}}{\partial_z p_0}\right)_{\textrm{Smol}} = \frac{\zeta_{|\textrm{w}} \lambda^2_{\textrm{D}}}{\eta_\infty D_{\infty,\textrm{n}}}.
\end{equation}
On the other hand, in the same limit, the shear-induced streaming potential at $\partial_z p_0 = 0$ behaves according to $(\partial_z \phi_{\textrm{st}}/\partial_z \sigma)_{H \gg \lambda_{\textrm{D}}}\sim$ $(H/\lambda_{\textrm{D}})^{-1}$, i.e. it vanishes for a film thickness much larger than $\lambda_{\textrm{D}}$.

\begin{figure}
\center
	\includegraphics[width = 8.4cm]{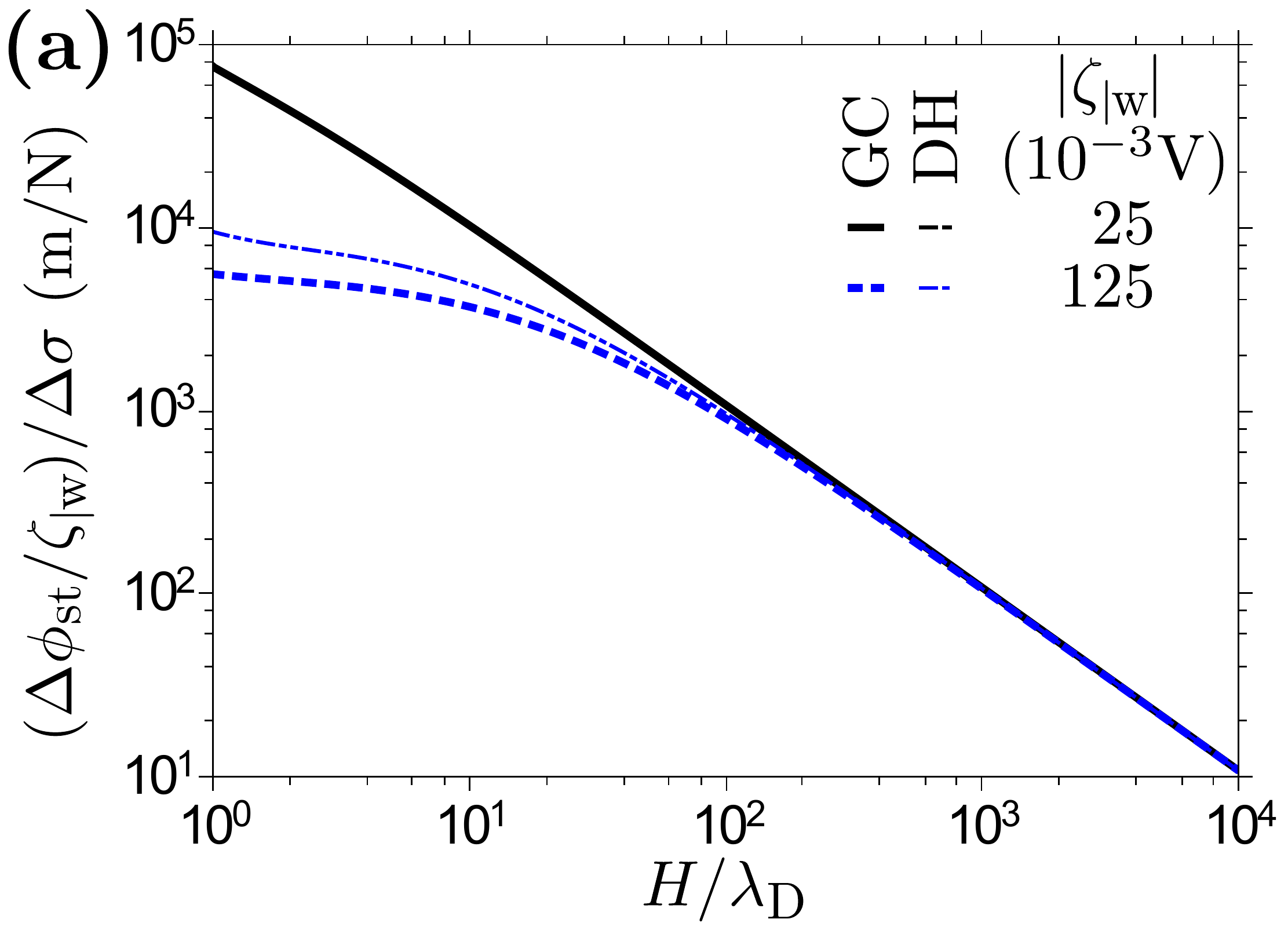}
	\includegraphics[width = 8.4 cm]{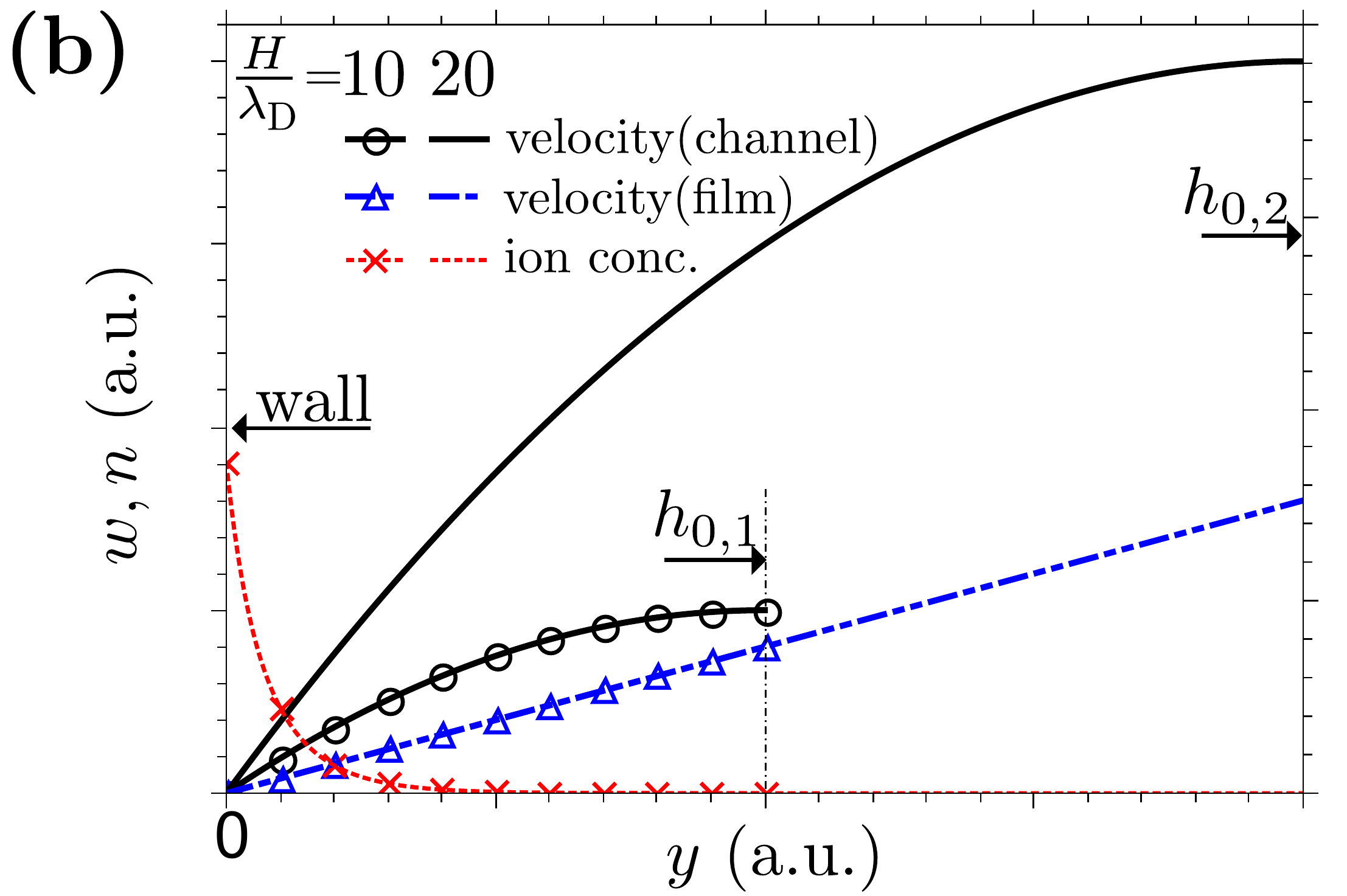}
	\includegraphics[width = 8.4 cm]{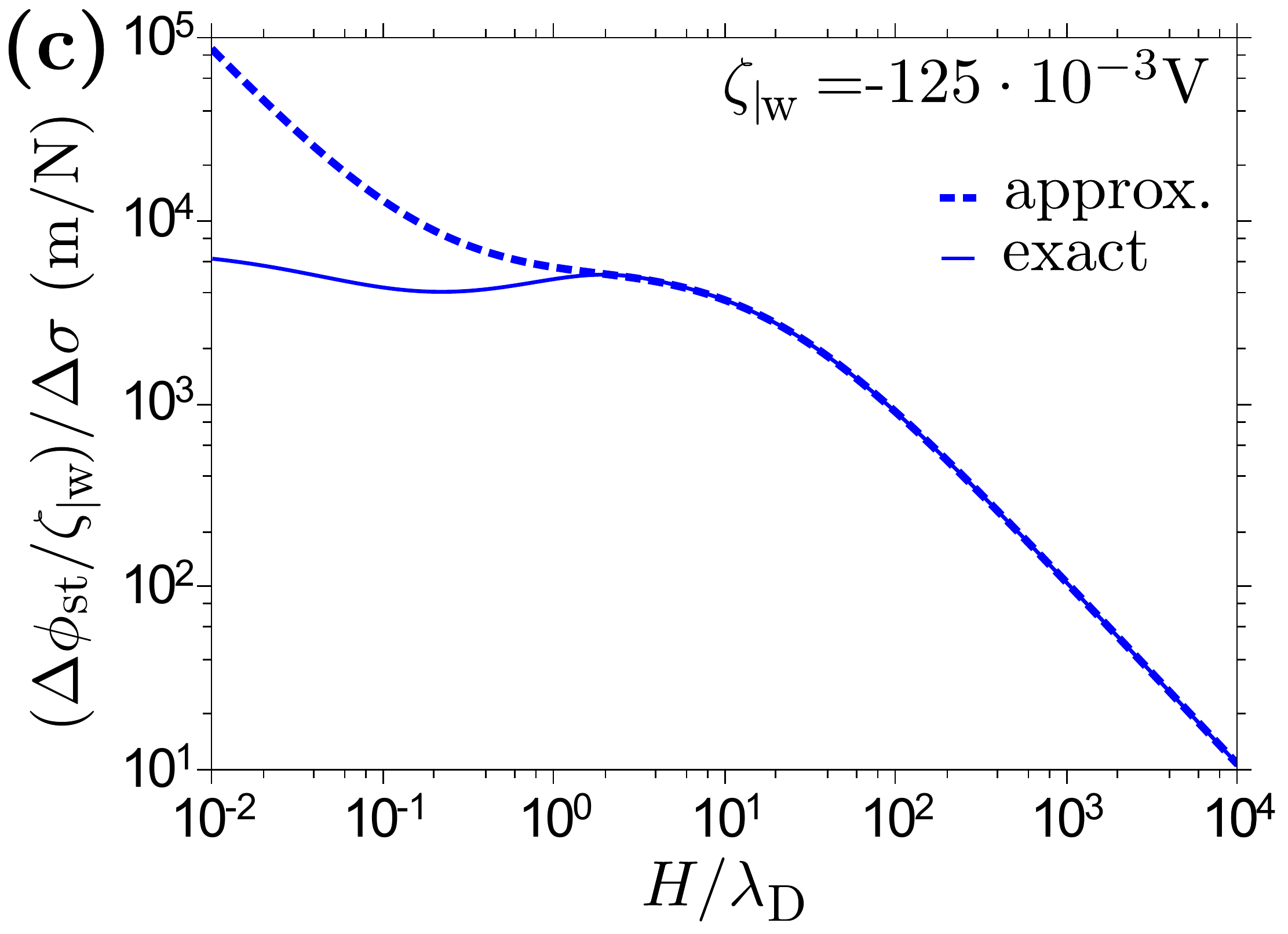}
	\caption{(a) Shear-induced streaming potential according to expression (\ref{Eq:streampot_noslip_nofield}), calculated either 
	with the Gouy-Chapman model (GC)(no EDL overlap but arbitrary $\zeta$-potentials) or with Debye-H\"uckel (DH) approximation 
	(small $\zeta$-potentials). The wall $\zeta$-potential is either $|\zeta_{|\textrm{w}}| = 25 \cdot 10^{-3} \textrm{V}$ or 
	$|\zeta_{|\textrm{w}}| = 125 \cdot 10^{-3}\:\textrm{V}$, while $\lambda_{\textrm{D}}\!\approx\! 10^{-7}\:\textrm{m}$. 
	(b) Schematic visualization of axial velocity distributions $w$ in pressure- or shear-driven flow, as well as 
	corresponding number concentrations $n$ of dominant ion species in EDL for either $h_{0,1} = 10\:\lambda_{\textrm{D}}$ or 
	$h_{0,2} = 20\:\lambda_{\textrm{D}}$. (c) Streaming potential calculated with (\ref{Eq:streampot}) (labeled 'exact') instead of 
	(\ref{Eq:streampot_noslip_nofield}) (labeled 'approx') if electric field at free surface left unscreened by EDL is considered.}
	\label{Fig:streampot_pershear}
\end{figure}

The behavior of $\partial_z \phi_{\textrm{st}}/\partial_z \sigma$ as a function of $H$ is depicted in figure \ref{Fig:streampot_pershear} (a). Results obtained with the Gouy-Chapman model (GC) are compared with approximations in the DH-limit, where in the latter case the corresponding EDL-potential is given by (\ref{Eq:EDLpot_DH_eqGC}). Two wall $\zeta$-potentials were used, either $|\zeta_{|\textrm{w}}| = 25 \cdot 10^{-3}\:\textrm{V}$ or $|\zeta_{|\textrm{w}}| = 125 \cdot 10^{-3}\:\textrm{V}$, while $\lambda_{\textrm{D}} \approx 10^{-7}\:\textrm{m}$. At the low $\zeta$-value, agreement between the GC- and the DH-model is excellent so that corresponding solutions completely overlap. For $|\zeta_{|\textrm{w}}| = 125 \cdot 10^{-3} \textrm{V}$, along with $H/\lambda_{\textrm{D}} \lesssim 100$, the DH-approximation overestimates the streaming potential. For $H/\lambda_{\textrm{D}} \lesssim 100$ alone, the streaming potential relative to the applied shear and $\zeta_{|\textrm{w}}$ (as plotted in (a)) is larger for smaller values of $\zeta_{|\textrm{w}}$ than for larger ones. The electro-osmotic and conductive backflow of the ions (second part in the denominator of (\ref{Eq:streampot_noslip_nofield}), driven by the streaming potential itself) increases quadratically with $|\zeta_{w}|$, reducing the streaming potential at higher $\zeta$-potentials. As expected, for $H \gg \lambda_{\textrm{D}}$, $(\partial_z \phi_{\textrm{st}}/\partial_z \sigma)_{H \gg \lambda_{\textrm{D}}} \sim (H/\lambda_{\textrm{D}})^{-1}$.

At $H \gg \lambda_{\textrm{D}}$, the vanishing streaming potential in shear-driven flow in comparison with the constant value obtained in pressure-driven flow can be explained by comparing the changes in the respective axial velocity profiles of these flow types upon an increase of $H$. This is schematically shown in figure \ref{Fig:streampot_pershear} (b). The abscissa contains the distance from the wall in an (for the present purpose of qualitative explanation) arbitrary length unit. The wall potential is equal to $\zeta_{|\textrm{w}}$. Also in arbitrary units, the ordinate shows the axial velocities or the number concentration of the dominant ion species in the EDL, respectively. As illustrated, the EDL-thickness remains practically unaffected if the film height $H > \lambda_{\textrm{D}}$ is increased from 10 to 20 multiples of $\lambda_{\textrm{D}}$. For pressure-driven flow, the peak velocity at $y = H$ increases quadratically with $H$ so that also the velocities within the EDL increase correspondingly, enhancing the convective ion transport. By contrast, for shear-induced flow, the flow velocities do not change within the EDL and the convective ion transport remains the same. For both types of flows, the conduction current increases linearly with $H$. Hence, pressure-driven flow compensates for the enhanced conduction current with increased $H$, whereas shear-driven flow does not. As a consequence, the shear-induced streaming potential vanishes for $H \gg \lambda_{\textrm{D}}$.

The use of the specific form of the integral expression in (\ref{Eq:streampot_noslip_nofield}) (which can be traced back to the simplified form of $F_{\textrm{CS}}$ expressed by (\ref{Eq:condfct2})) implies that effects caused by an incomplete screening of the wall charge by the EDL (e.g. in case of the film being very thin) is neglected. If considered, only the magnitude of the electro-osmotic and conductive backflow of ions is affected so that, at most, this is only relevant at higher values of $\zeta_{|\textrm{w}}$. For instance, in figure \ref{Fig:streampot_pershear} (c), the streaming potential as a function of $H/\lambda_{\textrm{D}}$ is depicted for $\zeta_{|\textrm{w}} = -125 \cdot 10^{-3}\:\textrm{V}$ when the electric field at the interface left unscreened by the EDL is calculated from the GC-model and used in the complete equation (\ref{Eq:streampot}). In other words, $E_{|\textrm{i}}$ is imposed in such a fashion that it corresponds to the value of $\zeta_{i}$ calculated from the GC-model. In this case, the latter is the accurate description of the EDL-potential even if the film is thinner than the EDL. This case is labeled with 'exact'. For comparison, the streaming potential calculated with (\ref{Eq:streampot_noslip_nofield}) is shown as well (labeled with 'approx'), which is identical to the corresponding case displayed in \ref{Fig:streampot_pershear} (a). As can be seen, the difference becomes visible only for film heights smaller than $\lambda_{\textrm{D}}$. Furthermore, the discrepancy is significantly reduced for lower values of $\zeta_{|\textrm{w}}$ (not shown).

\subsubsection{Shear-driven streaming potential with induced interfacial charge and wall slip}\label{Subsubsec:ifcharge_wallslip}

If an electric field $E_{|\textrm{i}}$ is used to induce a surface charge density $q_{|\textrm{i}}$ at the free interface, in the limit of large $H$ and vanishing slip length $b$, the shear-induced streaming potential reads
\begin{align}
\label{Eq:streampot_largeh0_iffield}
-\left(\frac{\partial_z \phi_{\textrm{st}}}{\partial_z \sigma}\right)_{E_{|\textrm{i}} \neq 0, H \gg \lambda_{\textrm{D}}} 
&= -\frac{E_{|\textrm{i}}}{\eta_\infty D_{\infty,\textrm{n}}/\lambda^2_{\textrm{D}} + \epsilon E^2_{|\textrm{i}}} \nonumber \\
&= \frac{q_{|\textrm{i}}}{\epsilon \eta_\infty D_{\infty,\textrm{n}}/\lambda^2_{\textrm{D}} + q^2_{|\textrm{i}}}.
\end{align}
In the following, this expression is abbreviated with $(\Delta \phi_{\textrm{st}}/\Delta \sigma)_{E_{|\textrm{i}},\infty}$.

\begin{figure}
\center
	\includegraphics[width = 8.4 cm]{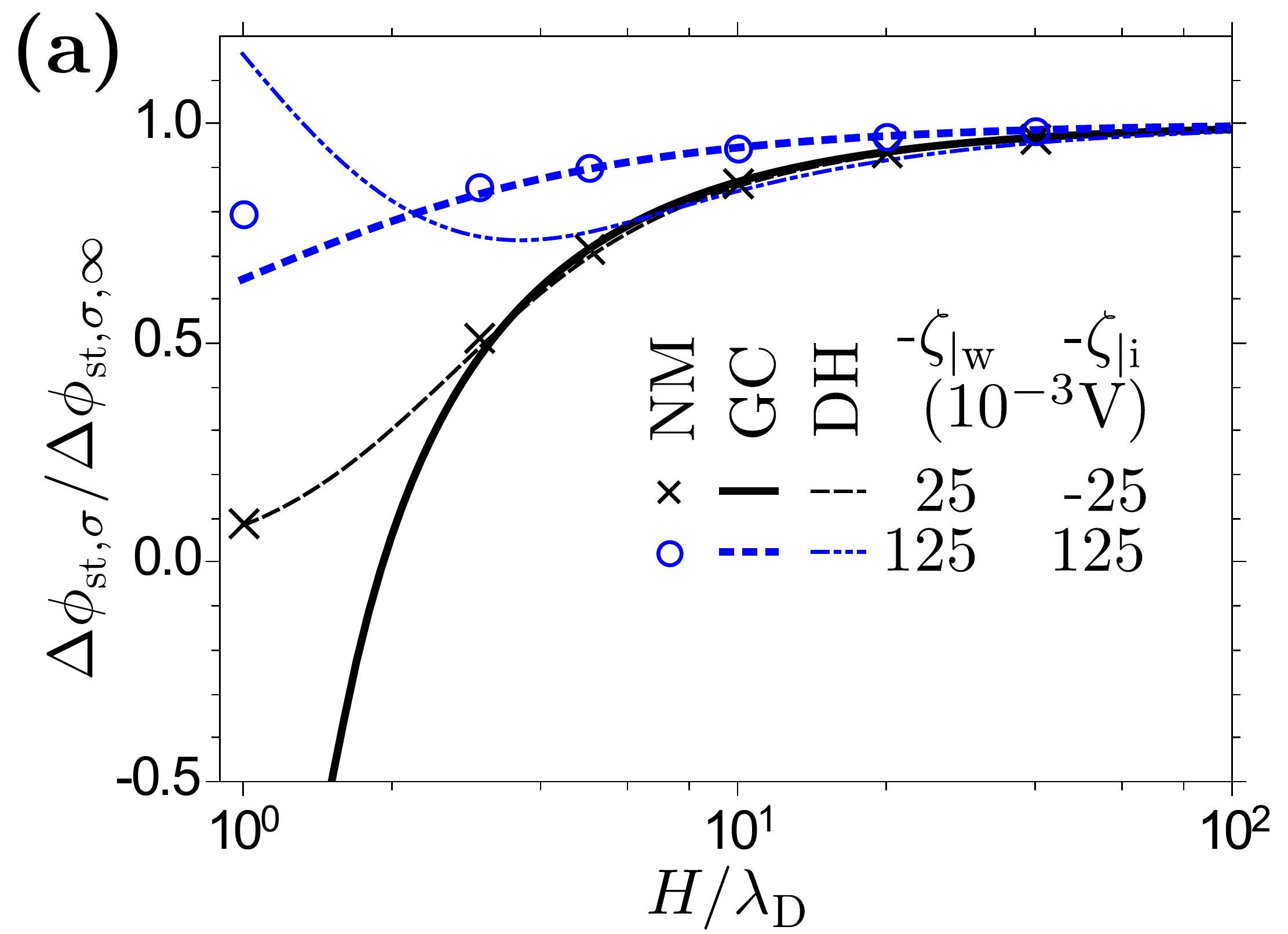}
	\includegraphics[width = 8.4 cm]{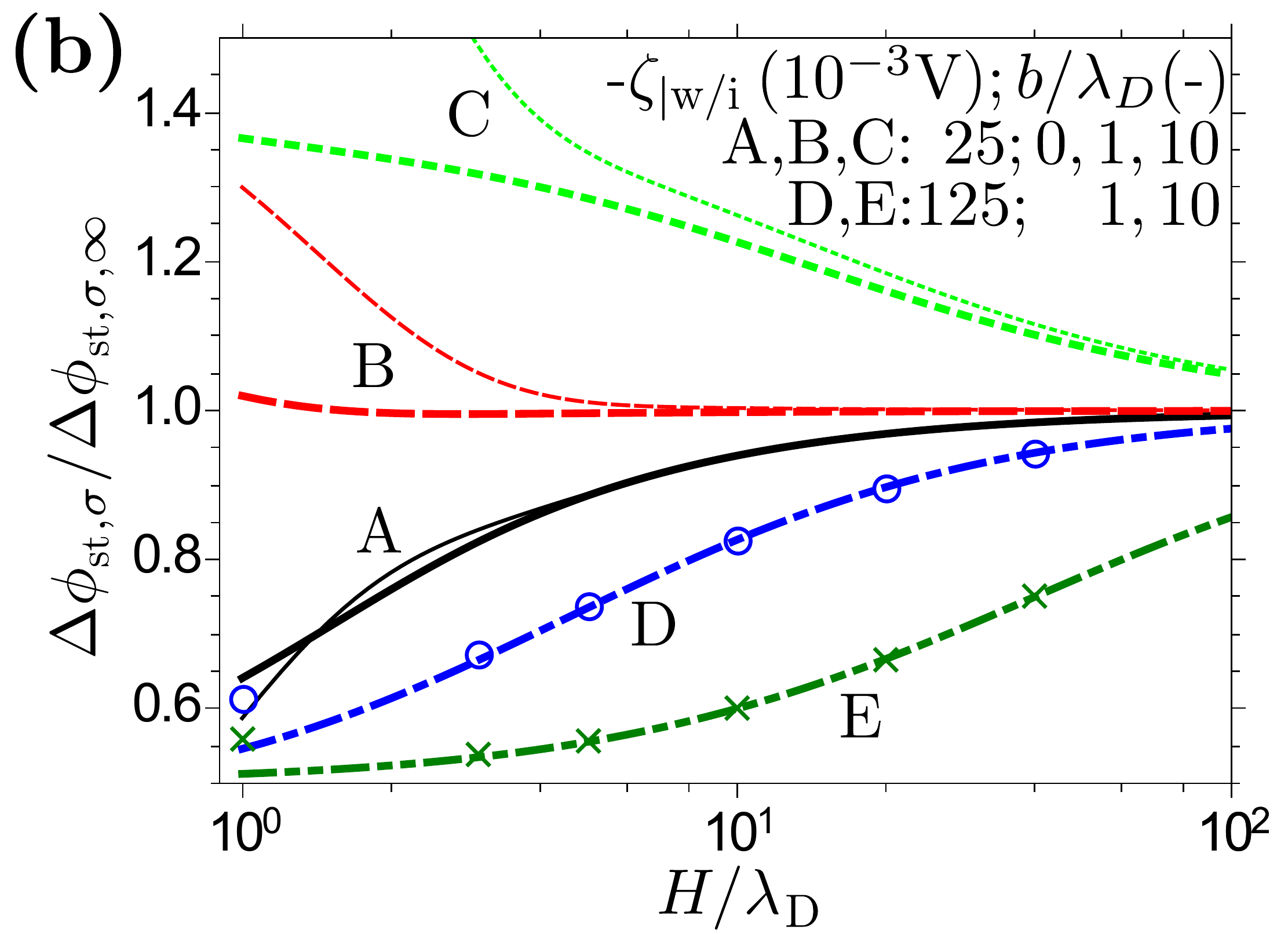}
	\caption{(a) Shear-induced streaming potential if the surface charge density is induced at the interface by an electric field $E_{|\textrm{i}}$, causing an interfacial potential $\zeta_{|\textrm{i}}$. Slip is excluded ($b\!=\! 0$). The local streaming potential $(\partial_z \phi_{\textrm{st}}/\partial_z \sigma)_{E_{|\textrm{i}} \neq 0}$, as determined by (\ref{Eq:streampot}) ($\partial_z p_0\!=\! 0$) is non-dimensionlized with its value for very large $H$, expressed by (\ref{Eq:streampot_largeh0_iffield}). For shorter notation, $(\Delta \phi_{st,\sigma})/(\Delta \phi_{st,\sigma,\infty}) = (\partial_z \phi_{\textrm{st}}/\partial_z \sigma)_{E_{|\textrm{i}}}/(\partial_z \phi_{\textrm{st}}/\partial_z \sigma)_{E_{|\textrm{i}} \neq 0,H \gg \lambda_{\textrm{D}}}$. Results from the Gouy-Chapman (GC) model are plotted in thick lines, while corresponding results obtained from the Debye-H\"uckel (DH) approximation are plotted in thin lines. Results from full numerical simulations (NM) are shown with symbols. The $\zeta$-potential pairs are either $(\zeta_{|\textrm{w}},\zeta_{|\textrm{i}}) = (-25,25) \times 10^{-3}\:\textrm{V}$ or $(\zeta_{|\textrm{w}},\zeta_{|\textrm{i}}) = (-125,-125) \times 10^{-3}\:\textrm{V}$. (b) Plot of $\Delta \phi_{st,\sigma}/(\Delta \phi_{st,\sigma,\infty})$ as a function of molecular wall slip length $b$ for either $(\zeta_{|\textrm{w}},\zeta_{|\textrm{i}}) = (-25,-25) \times 10^{-3}\:\textrm{V}$ (cases A-C) or $(\zeta_{|\textrm{w}},\zeta_{|\textrm{i}}) = (-125,-125) \times 10^{-3}\:\textrm{V}$ (cases D,E). The slip length is either zero (A, for reference), $b = \lambda_{\textrm{D}}$ (B,D) or $b\!=\! 10 \lambda_{\textrm{D}}$ (C,E), while $\lambda_{\textrm{D}} \approx 10^{-7}\:\textrm{m}$. At low $\zeta$-potential, the GC-model (thick lines) is compared with results obtained within the DH-approximation (thin lines), which fully agree with numerical simulation (not shown). At large $\zeta$-values, the GC-model is compared to results obtained by full numerical simulation (symbols).}
	\label{Fig:streampot_pershear_fieldandslip}
\end{figure}

In figure \ref{Fig:streampot_pershear_fieldandslip} (a), the shear-induced streaming potential is plotted as a function of $H/\lambda_{\textrm{D}}$ for the $\zeta$-potential pairs $(\zeta_{|\textrm{w}},\zeta_{|\textrm{i}}) = (-25,25) \times 10^{-3}\:\textrm{V}$ and $(-125,-125) \times 10^{-3}\:\textrm{V}$, respectively. Slip is not included ($b = 0$) so that the electric field at the wall itself (given by $E_{|\textrm{w}}$) has no direct effect on the streaming potential (other than being related to $\zeta_{|\textrm{w}}$). By contrast, the different values of the interfacial potentials 
$\zeta_{|\textrm{i}} = [25,-125] \times 10^{-3}\:\textrm{V}$ are induced by corresponding fields $E_{|\textrm{i}} \approx [-0.27,3.0] \times 10^6\:\textrm{V m}^{-1}$ (when $\lambda_{\textrm{D}}\!\approx\! 10^{-7}\:\textrm{m}$). Thus, expression (\ref{Eq:streampot_largeh0_iffield}) differs from zero and is used for non-dimensionlization of the (shear-induced) streaming potential $\Delta \phi_{\textrm{st}}/\Delta \sigma$ as a function of $H/\lambda_{\textrm{D}}$. As before, predictions according to the GC-model are compared with those obtained from the DH-solution. For the latter, the electric fields at the boundaries are given by (\ref{Eq:field_i_DH}) and (\ref{Eq:field_w_DH}), respectively, instead of expression (\ref{Eq:zeta_surfcharge}) used in the GC-model. Hence, solutions obtained at identical $\zeta$-potentials either from the GC-model or from the DH-model do not necessarily represent equal surface charge densities. For verification, solutions obtained by numerically solving the Poisson-Boltzmann equation (\ref{Eq:PBeq}) as well as the integral $\int^H_0 (\partial_y \psi/\psi_{\textrm{D}})^2 dy$ are plotted with symbols for selected values of $H/\lambda_{\textrm{D}}$. The boundary value problem was solved with the BVP4C-function implemented in Matlab R2012b, while the numerical integration was conducted employing the TRAPZ-function. For small $\zeta$-potentials and $H/\lambda_{\textrm{D}} \gtrsim 2.5$, good agreement between all three solution approaches is found, whereas the GC-model underpredicts the streaming potential for smaller values of $H/\lambda_{\textrm{D}}$. The deviation of the GC-model from the full numerical solution improves for $\zeta_{|\textrm{w}} = \zeta_{|\textrm{i}} = -125 \cdot 10^{-3}\:\textrm{V}$. However, for these larger $\zeta$-potentials, the DH-model first overpredicts ($H/\lambda_{\textrm{D}} < 2$), then underpredicts ($H/\lambda_{\textrm{D}} \gtrsim 2$) the streaming potential by up to $40 \%$ and only approaches the correct value for $H/\lambda_{\textrm{D}} > 30$. 

In the previous discussion of figure \ref{Fig:streampot_pershear} (a), it was highlighted that especially for smaller values of $H/\lambda_{\textrm{D}}$ and larger $\zeta$-potentials the electro-osmotic and conductive backflow of ions has a diminishing effect on the streaming potential. Owing to the third term in (\ref{Eq:denominator}), the corresponding effect is more involved if a surface charge is induced at the free surface. This is because this particular term of $F_{\phi}$ can either increase or decrease the streaming potential relative to $(\Delta \phi_{\textrm{st}}/\Delta \sigma)_{E_{|\textrm{i}},\infty}$, whereas all other terms contributing to $F_{\phi}$ always diminish it. For instance, for $\zeta_{|\textrm{i}} < 0$ (i.e. $E_{|\textrm{i}} > 0$) while $\zeta_{|\textrm{w}} - \zeta_{|\textrm{i}} > 0$, the streaming potential is enhanced. On the other hand it is reduced when $\zeta_{|\textrm{i}} > 0$ (i.e. $E_{|\textrm{i}} < 0$) while still $\zeta_{|\textrm{w}} - \zeta_{|\textrm{i}} > 0$. The possible parameter combinations are too manifold to be discussed exhaustively within the scope of the present work. In addition, such effects become only relevant for very small values of $H/\lambda_{\textrm{D}} \approx {\cal O}(1)$. For large $H/\lambda_{\textrm{D}}$, this contribution vanishes in all cases.

Figure \ref{Fig:streampot_pershear_fieldandslip} (b) illustrates the effect which a molecular slip length $b$ of ${\cal O}(\lambda_{\textrm{D}})$ has on $(\Delta \phi_{\textrm{st}}/\Delta \sigma)_{E_{|\textrm{i}}}$. Corresponding modifications of the shear-induced streaming potential are in effect only if $H$ is not much larger than $\lambda_{\textrm{D}}$. Hence, for $H \gg \lambda_{\textrm{D}}$, (\ref{Eq:streampot_largeh0_iffield}) remains valid even if the liquid molecules at the solid wall do slip for a distance $b \lesssim {\cal O}(\lambda_{\textrm{D}})$. In (b), two $\zeta$-potential pairs are used, either $(\zeta_{|\textrm{w}},\zeta_{|\textrm{i}}) = (-25,-25) \times 10^{-3}\:\textrm{V}$ (cases A-C) or $(\zeta_{|\textrm{w}},\zeta_{|\textrm{i}}) = (-125,-125) \times 10^{-3}\:\textrm{V}$ 
(cases D,E). The slip length is either zero (A, for reference), $b = \lambda_{\textrm{D}}$ (B,D) or $b = 10 \lambda_{\textrm{D}}$ (C,E). The results obtained with the GC-model are plotted with thick lines of different styles. At low $\zeta$-potentials, the GC-model is compared to corresponding predictions according to the DH-approximation plotted in the same style but with thin lines, which fully agree with numerical simulations using the full Poisson-Boltzmann equation (\ref{Eq:PBeq}) (not shown). If slip is included at low $\zeta$-potentials and small $H/\lambda_{\textrm{D}}$, the GC-model significantly underpredicts the achievable streaming potential. This indicates that the consideration of EDL-overlap is of crucial importance in these cases. At large $\zeta$-potentials, the GC-model is compared (at selected values of $H/\lambda_{\textrm{D}}$) with results obtained from numerical simulations, which are denoted by symbols. As apparent from this plot, molecular slip increases the streaming potential with increasing slip length only if, next to $H \lesssim b$, the $\zeta$-values are sufficiently low. For larger $\zeta$-potentials, the fifth term in (\ref{Eq:denominator}), quantifying the slip-enhancement of the electro-osmotically driven counter flow, grows faster with $b$ than the slip-induced enhancement of the convective ion transport, as expressed by the nominator of (\ref{Eq:streampot}). Alternatively, for large $H/\lambda_{\textrm{D}}$, this also follows directly from (\ref{Eq:streampot_largeh0_general}). Hence, for larger $\zeta$-potentials, the (shear-induced) streaming potential is decreased with increasing slip length.

In the next section, in the limit of infinitely thin EDL, an example of a technically feasible approach of shear-induced ion streaming is discussed. In this setting we also turn to the achievable efficiencies for energy conversion on the backdrop of surface-driven flow combined with large (apparent) slip.

\section{Electrokinetic streaming in shear-driven channel flow with superhydrophobic surfaces}\label{Sec:channel}

\begin{figure}
\centering
\includegraphics[width=8.4 cm]{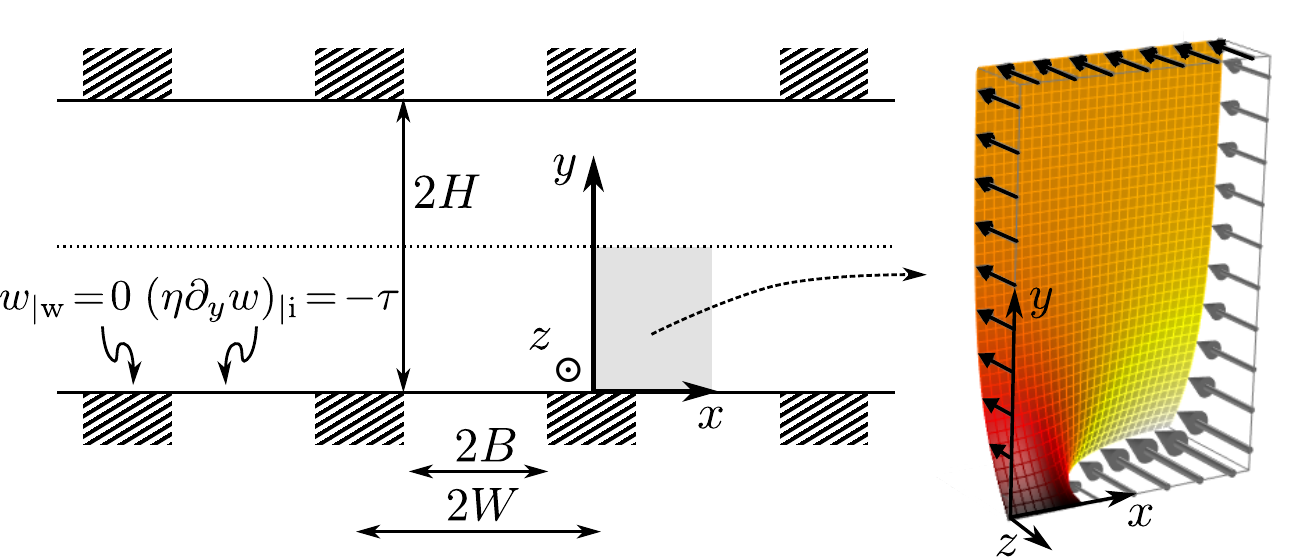}
\caption{Sketch of the geometry considered. 
Fluid is in Cassie-Baxter state between two structured plates at distance $2H$. The striped areas are posts with no-slip walls, between them is a planar liquid-gas interface where a constant stress $\eta(\partial_y w)_{|\textrm{i}} = -\tau$ is exerted. Due to symmetry it is sufficient to solve the Laplace equation, $\nabla^2 w=0$, in the gray area. An exemplary solution of the flow field within this unit cell is shown on the right (obtained with Comsol Multiphysics).
\label{Fig:schematic_microstructure}}
\end{figure}

The setup under discussion is schematically shown in figure~\ref{Fig:schematic_microstructure}. A symmetric electrolyte flows in a stationary fashion across a micro-structured surface of length $S$, where the periodicity is given by $2W$. The distance between the ribs is $2B$ so that the free surface fraction can be defined by $a = B/W$. All dimensions of the micro-structure are assumed to be much larger than the Debye length $\lambda_{\textrm{D}}$. The electrolyte is assumed to be in the Cassie-Baxter state, implying that it does not enter into the surface grooves \cite{Oh:EurophysLett2011}. Therefore, the electrolyte domain is bounded by a liquid-solid interface and a liquid-gas interface, which is assumed to remain flat. Given the large difference in viscosity between the liquid and the gas \cite{Schoenecker:JFM2014}, the shear stress within the gas phase trapped inside the surface grooves will be neglected.

The chosen configuration is an idealization and a special case in the sense that in an experimental realization the fluid may at least partially enter the grooves. Furthermore, for a given free surface fraction one would employ a periodic pattern of pillars rather than ribs simply to give more stability to the capillary surface. However, the chosen setup allows for an analytical treatment and permits to highlight the main physical effects clearer and more intuitively than possible at hand of full numerical treatments. Furthermore and more importantly, we are specifically interested in an upper limit of the conversion efficiency from mechanical to electric energy, for which the chosen idealization is particularly helpful.

The flow is induced by a shear stress $\partial_z \sigma$ which acts along the free interface between the ribs in longitudinal direction. The solid surface in contact with the electrolyte is charged and causes the accumulation of an ion cloud in the electrolyte of surface charge density $q_{|\textrm{w}}$. An external electric field is applied perpendicular to the spread direction of the free surface, which induces a charge density $q_{|\textrm{i}}$. The electric field $E_{|\textrm{i}}$ used to induce $q_{|\textrm{i}}$ is limited by the electrohydrodynamic stability of the interface and the break-down voltage of the surrounding air \cite{Oh:EurophysLett2011}. The former provides an upper estimate for the spacing $2B$, while the latter limits $E_{|\textrm{i}}$ to $O(10^6\: \textrm{V m}^{-1})$. As shown in the previous sections, since all of the geometric parameters are assumed to substantially exceed $\lambda_{\textrm{D}}$, $q_{|\textrm{w}}$ can be expected to have a negligible effect on the electrokinetic streaming. The flow is assumed to be fully developed so that the non-linear part of the Navier-Stokes equations can be neglected. Under steady-state, the flow is then governed by the Stokes equations, where the axial velocity is described by (\ref{Eq:NSEax}).

\subsection{Velocity} \label{Sec:vel_mar}

The shear stress along the free surface leads to an axial fluid velocity $w_\sigma$. In turn, the electric net charge convectively transported with this flow causes a charge polarization and a corresponding (induced) streaming field $\partial_z \phi_{\textrm{st}}$. Next to the conduction current in the bulk, this gradient in electric potential drives an overall electro-osmotic counter flow, denoted herein by $w_{q_{|\textrm{s}}}$. In this context, the fluid velocity due to the electro-osmotic fluid propulsion related to the presence of the ion cloud at the solid wall is denoted by $w_{q_{|\textrm{w}}}$, and $w_{q_{|\textrm{i}}}$ describes the corresponding flow due to the ion cloud at the free surface. Given the linearity of the Stokes equation, all of these velocity contributions can be treated separately. The total axial fluid velocity $w$ is then given by their linear superposition. In the following, for notational simplicity, $\eta \equiv \eta_\infty$. 

Firstly, the flow $w_\sigma$ due to a surface tension gradient, inducing a constant shear stress $\eta(\partial_y w_y)_{|\textrm{i}} = -\partial_z \sigma$ at the gas-liquid interface, is considered. On the solid wall the no-slip condition applies, while due to symmetry the shear rate vanishes in the channel center-plane, $(\partial_y w)_{|y=H} = 0$. As long as no back pressure is applied, equation (\ref{Eq:NSEax}) reduces to the Laplace equation, $(\partial^2_x+\partial^2_y) w_\sigma = 0$, for the velocity in $z$-direction. 

In the limit of infinite separation between the plates, Philip \cite{Philip:ZAngewMath1972} (case 5) showed that the solution for $w_\sigma$ above the lower surface is given by
\begin{equation} \label{Eq:philip_sigma}
w_{\sigma}(x,y) = -\frac{W \partial_z \sigma}{\eta} \left[\Phi_0(x,y) - \frac{y}{W}\right].
\end{equation}
Here
\begin{equation} \label{Eq:non_dim_vel_field}
\Phi_0(x,y) = \frac{2}{\pi} \Im \left[\arcsin \left(\frac{\sin(\frac{\pi}{2 W}(x+ i y))}{\sin(\frac{\pi}{2}(1-a))} \right) \right] 
\end{equation}
is a non-dimensional velocity field with the properties $\Phi_0(x,0) = 0$ for $0<x<(1-a)W$ (i.e. at the solid wall) and $\partial_y\Phi_0(x,0) = 0$ for $(1-a)W<x<W$ (i.e. at the free surface). Far away from the surface, $\lim_{y\to\infty}\partial_y\Phi_0=W^{-1}$. $\Im(\xi)$ is the imaginary part of the complex number $\xi$ and $i$ denotes the imaginary unit. Strictly speaking, (\ref{Eq:philip_sigma}) (with (\ref{Eq:non_dim_vel_field})) is only a valid solution for the present problem in the limit $H\to\infty$. However, $w_\sigma(x,y)$ rapidly approaches a constant for $y\gg W$, in particular
\begin{equation}\label{Eq:philip_asymptotic}
[\Phi_0(x,y) - y/W] - \beta_{\|} \sim e^{-\pi y/W},
\end{equation}
where
\begin{equation}\label{Eq:beta}
\beta_{\|} = -\frac{2}{\pi}\ln \cos\left( \frac{\pi a}{2}\right)
\end{equation}
is the non-dimensional velocity scale for this type of flow. In consequence, the parameters relevant herein, i.e. the flow rate and the line-average of the velocity at $y=0$, are excellently approximated by (\ref{Eq:philip_sigma}) already for $H\gtrsim W$. A numerical quantification of this assertion is made in appendix \ref{sec:appendix_checkAnalytic}, where it is also shown how flow rates and line averages of the velocity are related.

Next, the velocity field $w_{q_{|\textrm{i}}}$ (originating from the charge accumulated at the gas-liquid interface) is described. In the limit of an infinitely thin EDL as treated herein, the overall electro-osmotic force per volume on the ion cloud accumulating in the vicinity of the free surface can be replaced \cite{Gao:ColSuA2005,Steffes:ColSuA2011} by an effective stress condition $\eta (\partial_y w_{q_{|\textrm{i}}})_{|\textrm{i}} = -q_{|\textrm{i}} \partial_z \phi_{\textrm{st}} = -\tau_{\infty,q}$, while in the bulk $(\partial^2_x + \partial^2_y) w_{q_{|\textrm{i}}} = 0$. All solid surfaces are subject to the no-slip condition. Together with the symmetry condition $(\partial_y w)_{|y=H} = 0$ on the center-line between the two plates one thus has the same scenario as treated when solving for $w_\sigma$. Hence, $w_{q_{|\textrm{i}}}$ is given by 
\begin{equation} \label{Eq:philip_qi}
w_{q_{|\textrm{i}}}(x,y) = -\frac{W q_{|\textrm{i}} \partial_z \phi_{\textrm{st}}}{\eta} \left[\Phi_0(x,y) - \frac{y}{W}\right].
\end{equation}

The electro-osmotically driven fluid velocity along a solid (no-slip) wall adjacent to an ion cloud screening a constant surface charge density $-q_{|\textrm{w}}$ is generally given by
\begin{equation} \label{Eq:EOF_solidwall_qw}
w_{q_{|\textrm{w}}}(y) = -\frac{\epsilon}{\eta} [\psi(y)-\zeta_{|\textrm{w}}] \partial_z \phi_{\textrm{st}}.
\end{equation}
Within the DH-approximation the potential within the EDL can be approximated by $\psi \approx \zeta_{|\textrm{w}} \textrm{exp}(-y/\lambda_{\textrm{D}})$, so that in close proximity of the no-slip wall one has
\begin{equation} \label{Eq:philip_qw}
w_{q_{|\textrm{w}}}(y) = u_\mathrm{HS} \left[\textrm{exp}(-y/\lambda_{\textrm{D}})-1\right]. 
\end{equation}
The Helmholtz-Smoluchowski (HS) velocity is denoted by $u_\mathrm{HS} = -\epsilon \zeta_{|\textrm{w}} \partial_z \phi_{\textrm{st}}/\eta$. In the limit of an infinitely thin EDL, $w_{q_{|\textrm{w}}} \approx -u_{HS}$. The wall $\zeta$-potential as a function of $q_{|\textrm{w}}$ is given by (\ref{Eq:zeta_surfcharge}) or can be approximated by $\zeta_{|\textrm{w}} \approx - q_{|\textrm{w}} \lambda_{\textrm{D}}/\epsilon$. Thus, for a finite wall charge, the HS-velocity vanishes in the limit $\lambda_{\textrm{D}}\to 0$; therefore, this contribution will be neglected from here on.

Note, however, that in the case of an uncharged free surface the electro-osmotic flow is dominated by the charge on the no-slip region and the velocity profile between the plates essentially constitutes a plug-flow of velocity $u_{HS}$ (apart from the region of size $\sim\lambda_{\textrm{D}}$ at the no-slip wall), since the free-slip surfaces do not contribute to viscous dissipation \cite{Squires:POF2008}. Also note that the HS-velocity scales as $u_\mathrm{HS}=\lambda_{\textrm{D}} q_{|\textrm{w}} (\partial_z \phi_{\textrm{st}}/\eta)$ while the electro-osmotic velocity due to charges in the gas-liquid interface scales as $(\beta_\| W) q_{|\textrm{i}} (\partial_z \phi_{\textrm{st}}/\eta)$, i.e. the relevant length scale defining the former is $\lambda_{\textrm{D}}$ while the latter scales with $\beta_\| W \gg \lambda_{\textrm{D}}$; this again confirms that the influence of $u_\mathrm{HS}$ can safely be neglected for our purposes.

\subsection{Flow rates and electric currents} \label{Sec:flow_curr}

The overall flow rate is determined by integrating the velocity distribution across half of the channel height, i.e. 
\begin{equation} \label{Eq:flow_rate_general}
\dot{Q} = W^{-1} \int^W_0\!\!\!{\int^H_0{w\:dy}\:dx},
\end{equation}
where $w=w_\sigma + w_{q_{|\textrm{i}}}$ as expressed in eqns. (\ref{Eq:philip_sigma}) and (\ref{Eq:philip_qi}). According to Philip \cite{Philip:ZAngewMath1972b} (or using the asymptotic behavior of (\ref{Eq:philip_asymptotic}) together with equation (\ref{Eq:lineIntegrals}) discussed in the appendix), 
\begin{equation}
\int_0^W\!\!\! \int_0^H (\Phi_0-y/W) \: dy \: dx=HW\beta_\|
\end{equation}
and the flow rate can be calculated to read
\begin{equation} \label{Eq:flow_rate_EOF}
\dot{Q} = - L_{Q,q} \partial_z \phi_{\textrm{st}} - L_{Q,\sigma} \partial_z \sigma,
\end{equation}
where
\begin{equation} \label{Eq:LQq}
L_{Q,q} = \frac{q_{|\textrm{i}}}{\eta} HW \beta_{\|}
\end{equation}
and
\begin{equation} \label{Eq:LQsigma}
L_{Q,\sigma} = \frac{1}{\eta} HW \beta_{\|},
\end{equation}
with $\beta_\|$ as dimensionless velocity scale defined by (\ref{Eq:beta}). 

The streaming current is determined by
\begin{align} \label{Eq:stream_curr_general}
I_{\textrm{st}} = W^{-1} &\int^W_0\!\!\!{\int^H_0{\rho_e w\:dy}\:dx} = -\frac{q_{|\textrm{i}} \partial_z \phi_{\textrm{st}} + \partial_z \sigma}{\eta} \times \nonumber \\
&\int^W_0{\mkern-12mu\int^H_0{\!\!\rho_e(x,y) [\Phi_0(x,y)-y/W] \,dy}\,dx}.
\end{align}
In $y$-direction, $(\Phi_0-y/W)$ changes on the scale of $W$, whereas $\rho_e(x,y)$ varies within the length scale $\lambda_{\textrm{D}} \ll W$ and is zero outside the EDL. Hence, one can safely make the approximation
\begin{align} \label{Eq:stream_curr_general2}
I_{\textrm{st}} &\approx -\frac{q_{|\textrm{i}} \partial_z \phi_{\textrm{st}} + \partial_z \sigma}{\eta}\times\nonumber\\
&\mkern63mu \int^W_0{[\Phi_0(x,y)-y/W]\int^H_0{\rho_e(x,y)\,dy}\,dx} \nonumber \\
&\approx - L_{I,q} \partial_z \phi_{\textrm{st}} - L_{I,\sigma} \partial_z \sigma 
\end{align}
with
\begin{equation} \label{Eq:LIq}
L_{I,q} = \frac{q^2_{|\textrm{i}}}{\eta} W \beta_{\|}
\end{equation}
and
\begin{equation} \label{Eq:LIsigma}
L_{I,\sigma} = \frac{q_{|\textrm{i}}}{\eta} W \beta_{\|}.
\end{equation}
Note that within a distance $\lambda_{\textrm{D}}$ from the solid wall $\Phi_0-y/W \approx 0$, so that the transport of $q_{|\textrm{w}}$ by $w_{q_{|\textrm{i}}}$ can safely be neglected in our situation. In summary, the volumetric flux and the streaming current in the channel can be expressed by
\begin{equation} \label{Eq:Onsager_system}
\left(\begin{array}{c}
\dot{Q} \\
I_{\textrm{st}}
\end{array}\right)
= -
\left(\begin{array}{cc}
L_{Q,q} & L_{Q,\sigma} \\
L_{I,q} & L_{I,\sigma}
\end{array}\right)
\left(\begin{array}{c}
\partial_z \phi_{\textrm{st}} \\
\partial_z \sigma
\end{array}\right),
\end{equation}
where the $L_j$ are the Onsager coefficients as expressed above.

\subsection{Efficiency} \label{Sec:efficiency}

\begin{figure}
\center
\includegraphics[width=8.4 cm]{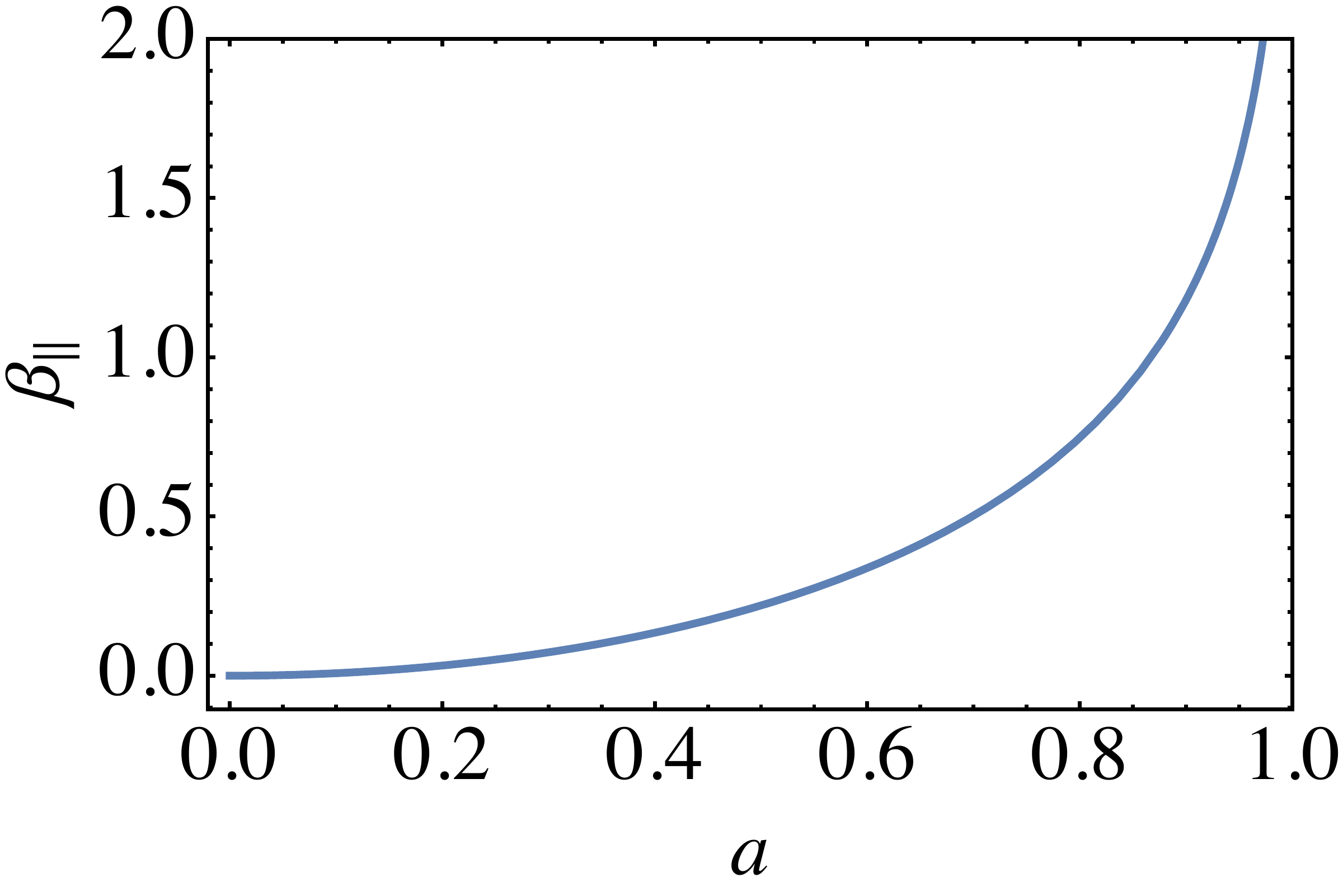}
\includegraphics[width=8.4 cm]{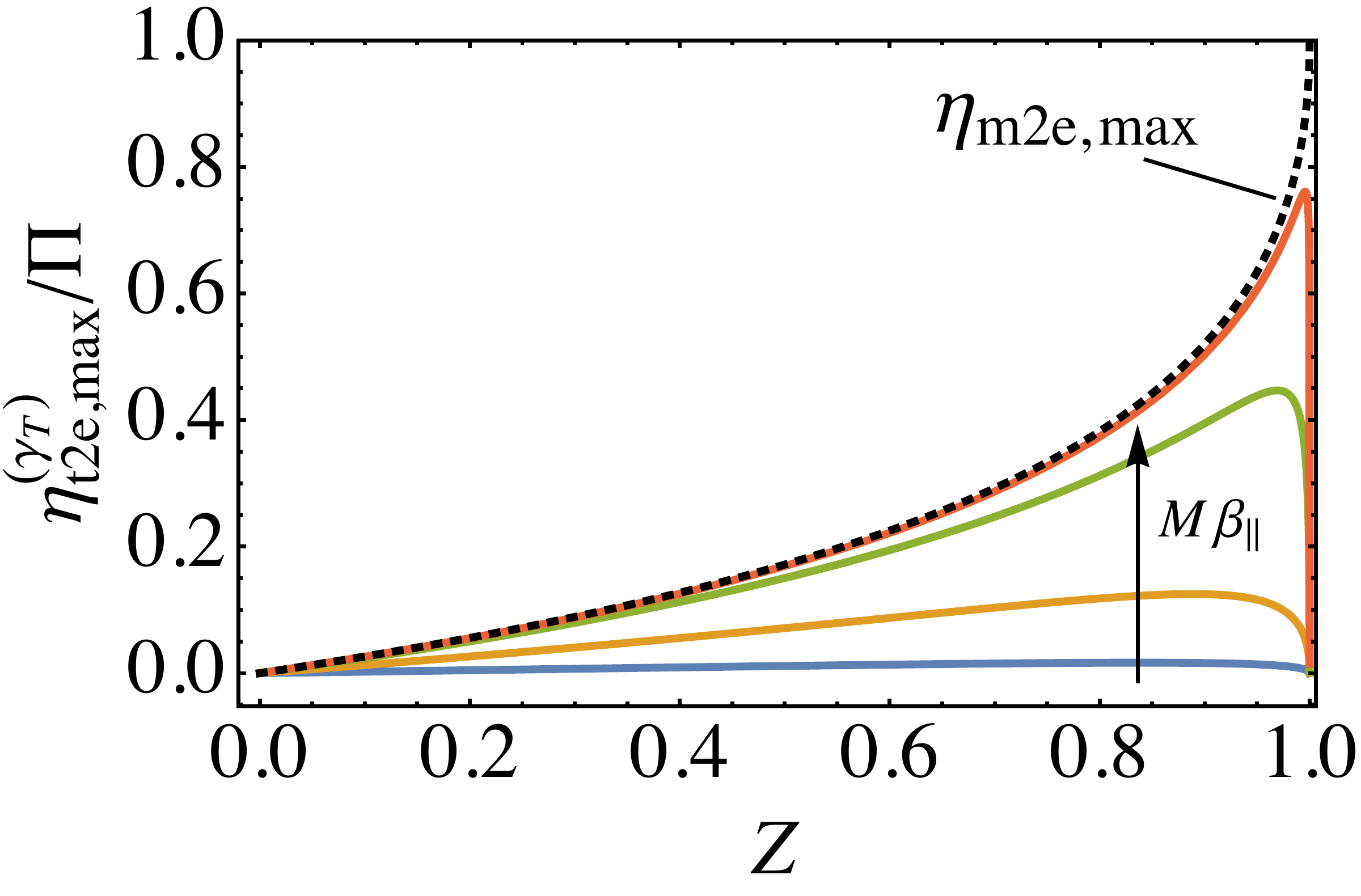}
\caption{(a) Dimensionless velocity scale $\beta_\|$ vs. free surface fraction $a$, according to eq. (\ref{Eq:beta}). Note that $\beta_\|$ diverges for $a\to 1$. This unphysical behavior is explained in \cite{Schoenecker:JFM2014}, along with a remedy. (b) Scaled thermal efficiency at optimal electric load, $\eta^{(\gamma_T)}_{\textrm{t}2\textrm{e},\textrm{max}}/\Pi$, vs. figure of merit $Z$, according to eq. (\ref{Eq:eff_t2e}). The product of the Marangoni number $M$ and $\beta_\|$ assumes values according to $M \beta_{\|}\!=\![0.1,1,10,100]$. In realistic situations $\beta_\|$ is typically of ${\cal O}(1)$ so that these values directly correlate to the Marangoni number. For $M \rightarrow \infty$, the conversion efficiency from mechanical to electric energy at optimal electric load, $\eta_{\textrm{m}2\textrm{e},\textrm{max}}$ (as expressed by (\ref{Eq:eff_m2e_press})), is recovered. This is indicated by the dashed line. From equation (\ref{Eq:figure_merit2}), it is evident that the figure of merit approaches 1 for a large velocity scale, $\beta_\|$, and is reduced by viscous dissipation and loss currents scaling with the viscosity, $\eta$, and conductivity, $\sigma^\infty$, respectively.}
\label{Fig:Beta_vs_a_Eta_vs_Z}
\end{figure}

Following (\ref{Eq:condcurr2}) with $F_{\textrm{CS}} \approx H$, the conduction current in the bulk can be approximated by
\begin{equation}\label{Eq:condcurr3} 
I_{\textrm{cd}} \approx -\epsilon\frac{D_{\textrm{n}} H}{\lambda^2_{\textrm{D}}}\partial_z \phi_{\textrm{st}} = -\sigma^{\infty} H \partial_z \phi_{\textrm{st}},
\end{equation}
with the bulk conductivity $\sigma^\infty$ defined in the paragraph following eq. (\ref{Eq:streampot_largeh0_general}). If used as an energy converter, the electrokinetic streaming device is embedded in a closed electric circuit with an external electric consumer of electric resistance\footnote{\label{FN:resistance} Since only half of the channel height is considered, the corresponding total external resistance for the full channel, i.e. two half-systems in parallel, is $R/2$.} $R$  aligned in parallel to the internal electric resistance of the energy converter itself. With $\Delta \phi_{\textrm{st}} = S \partial_z \phi_{\textrm{st}}$, overall charge conservation requires $-L_{I,q}\partial_z \phi_{\textrm{st}} - L_{I,\sigma} \partial_z \sigma = [\sigma^{\infty} H + S/(R W)]\partial_z \phi_{\textrm{st}}$ so that
\begin{equation}\label{Eq:charge_cons} 
-\partial_z \phi_{\textrm{st}} = \frac{L_{I,\sigma}}{L^{(0)}_{I,q} + S/(R W)} \partial_z \sigma,
\end{equation}
where $L^{(0)}_{I,q} = L_{I,q} + \sigma^{\infty} H$. If $R \rightarrow \infty$, one obtains the streaming potential under vanishing external load, namely
\begin{equation}\label{Eq:stream_pot_noload} 
-\left(\partial_z \phi_{\textrm{st}}\right)_{|R \rightarrow \infty} = \frac{q_{|\textrm{i}} \beta_{\|} W/H}{\eta \sigma^{\infty} + q^2_{|\textrm{i}} \beta_{\|} W/H} \partial_z \sigma.
\end{equation}
For $\partial_z p \equiv 0$ and identifying $(1+b/H) \hat{=} \beta_{\|} W/H$, this expression agrees with (\ref{Eq:streampot_largeh0_general}). If $R$ remains finite, the power extracted by the consumer reads
\begin{equation}\label{Eq:power_ex} 
P_{\textrm{ex}} = \frac{(\Delta \phi_{\textrm{st}})^2}{R} = \left[\frac{L_{I,\sigma}}{L^{(0)}_{I,q} + S/(R W)}\right]^2 \frac{(\Delta \sigma)^2}{R},
\end{equation}
where $\Delta \sigma = S \partial_z \sigma$. The mechanical power fed into the system equals
\begin{equation}\label{Eq:power_in} 
P_{\textrm{in}} = a W S \partial_z \sigma\, \overline{w}_{|\textrm{i}},
\end{equation}
where $\overline{w}_{|\textrm{i}}$ is the averaged axial velocity along the free surface. Given the no-slip condition 
at $y=0$ along the solid wall, one can write
\begin{align}\label{Eq:wi_avg} 
\overline{w}_{|\textrm{i}} &= \frac{1}{B} \int^W_{W-B}{w_{|\textrm{i}} dx} =  \frac{1}{a W} \int^W_0{\!\Phi_0(x,0)\! dx} \nonumber\\
&= -\frac{1}{a H}(L_{Q,q} \partial_z \phi_{\textrm{st}} + L_{Q,\sigma} \partial_z \sigma).
\end{align}
With (\ref{Eq:charge_cons}) the conversion efficiency from mechanical to electric energy thus reads
\begin{align}
\label{Eq:eff_m2e} 
\eta_{\textrm{m}2\textrm{e}} = \frac{P_{\textrm{ex}}}{P_{\textrm{in}}} = &\frac{S/(R W)}{L^{(0)}_{I,q} + S/(R W)} \times \nonumber \\
&\frac{H L^2_{I,\sigma}}{L_{Q,\sigma} [L^{(0)}_{I,q} + S/(R W)] - L_{Q,q} L_{I,\sigma}}
\end{align}
According to (\ref{Eq:LQq}) and (\ref{Eq:LIsigma}), one has $L_{Q,q} = H L_{I,\sigma}$. In addition, following the notation used by Xuan \textit{et al.} \cite{Xuan:JPowerSource2005} and Heyden \textit{et al.} \cite{vanderHeyden:NanoLett2006} with the dimensionless parameter
\begin{equation}\label{Eq:resitivity} 
\Theta = L^{(0)}_{I,q} R W/S
\end{equation}
and the (dimensionless) figure of merit
\begin{equation}\label{Eq:figure_merit} 
Z = \frac{L^2_{I,\sigma}}{L_{Q,\sigma} L^{(0)}_{I,q}}H
\end{equation}
one finds
\begin{equation}\label{Eq:eff_m2e2} 
\eta_{\textrm{m}2\textrm{e}} = \frac{Z \Theta}{(\Theta+1)(\Theta + 1 - Z \Theta)},
\end{equation}
which is formally identical to the conversion efficiency obtained for pressure-driven flow \cite{Xuan:JPowerSource2005}. Its maximum with respect to $\Theta$ (i.e. the external load)
\begin{equation}\label{Eq:eff_m2e_press} 
\eta_{\textrm{m}2\textrm{e},\textrm{max}} = \frac{Z}{(1+\sqrt{1-Z})^2} = \frac{(1-\sqrt{1-Z})^2}{Z}
\end{equation}
is reached for $\Theta_{|\textrm{max}} = 1/\sqrt{1-Z}$. Expression (\ref{Eq:eff_m2e_press}) is monotonously increasing with $Z$. Re-inserting the expressions for the Onsager coefficients, $Z$ can be expressed by
\begin{equation}\label{Eq:figure_merit2} 
Z^{-1} = 1+\frac{\eta \sigma^\infty}{q^2_{|\textrm{i}} \beta_{\|}} \frac{H}{W},
\end{equation}
i.e. $0<Z<1$, and $Z$ increases with $\beta_{\|}$. According to (\ref{Eq:beta}) and as shown in figure \ref{Fig:Beta_vs_a_Eta_vs_Z} (a), $\beta_{\|}$ monotonously increases with $a$ and hence so does $\eta_{\textrm{m}2\textrm{e},\textrm{max}}$. The latter is depicted as a function of $Z$ by the dashed line in figure \ref{Fig:Beta_vs_a_Eta_vs_Z} (b).

\subsubsection{Thermocapillarity-induced streaming}\label{Sec:thermocap}

In the following, the special case is considered that the mechanical surface stress is caused by a temperature-dependent surface tension, i.e $\partial_z \sigma = -\gamma_T \partial_z T$, where $\gamma_T$ is the change of surface tension with temperature $T$ (Marangoni coefficient). Disregarding the thermal conduction in the solid walls, the heating power required to maintain the axial temperature gradient reads
\begin{equation}\label{Eq:heat_flux}
P_{\textrm{th}} = -(\rho c_p S \overline{w} + k) H W \partial_z T,
\end{equation}
where $\rho$ is the density, $c_p$ the heat capacity and $k$ the thermal conductivity of the electrolyte. With the average fluid velocity $\overline{w}\!=\! 1/W\! \int^W_0\! w dx\!=\! a \overline{w}_{|\textrm{i}}$ and $\alpha\!=\! k/(\rho c_p)$ as the thermal diffusivity, the conversion efficiency from thermal to mechanical energy is given by
\begin{equation}\label{Eq:eff_therm}
\eta^{(\gamma_T)}_{\textrm{t}2\textrm{m}} = \frac{P_{\textrm{in}}}{P_{\textrm{th}}} = \frac{\Pi}{1+\alpha/(S \overline{w})},
\end{equation}
with $\Pi = \gamma_T/(\rho c_p H)$. In case of maximum efficiency ($\Theta_{|\textrm{max}} = 1/\sqrt{1-Z}$), one finds
\begin{equation}\label{Eq:eff_therm_conduct}
\frac{\alpha}{S \overline{w}} = \frac{1}{M \beta_{\|}} \frac{1}{\sqrt{1-Z}},
\end{equation}
where $M = \gamma_T \Delta T W/(\eta \alpha)$ is the thermal Marangoni number.  With (\ref{Eq:eff_m2e_press}), the total conversion efficiency from thermal to electric energy can be written as
\begin{equation}\label{Eq:eff_t2e} 
\eta^{(\gamma_T)}_{\textrm{t}2\textrm{e},\textrm{max}} = \frac{\Pi}{1+1/(M\beta_\| \sqrt{1-Z})} \frac{Z}{(1+\sqrt{1-Z})^2}
\end{equation}
This expression, scaled by $\Pi$, is plotted in figure \ref{Fig:Beta_vs_a_Eta_vs_Z} (b) for $M \beta_{\|}\!=\![0.1,1,10,100]$. In this plot, the dashed line indicates the limiting case of $M \rightarrow \infty$, for which the conversion efficiency from mechanical to electric energy at optimal electric load, $\eta_{m2e,max}$ (as expressed by (\ref{Eq:eff_m2e_press})), is recovered. At smaller values of $M$, the conversion efficiency from thermal to electric energy is at its maximum for
\begin{equation}\label{Eq:Zmax} 
Z_{|\textrm{max}} = 1\!-\! \frac{1}{3+2\left(M \beta_{\|}\!+\! \sqrt{2} \sqrt{1\!+\! M \beta_{\|}}\right)},
\end{equation}
where $Z_{|\textrm{max}} \rightarrow 1$ for large $M$.

The factor $\Pi$ appears to be the main limiting factor for $\eta^{(\gamma_T)}_{\textrm{t}2\textrm{e},\textrm{max}}$ since the latter cannot exceed this value even for $Z \rightarrow 1$ (e.g. if $q_{|\textrm{i}}$ is very high or by using discontinuous fluid domains so that $\sigma^{\infty}$ is vanishingly small). Electrolytes exhibiting a larger value of $\gamma_T$ than those using water as ion solvent are scarce. For the latter at $25\:^\circ \textrm{C}$, $\Pi H \approx 3.6 \cdot 10^{-11}\:\textrm{m}$. Hence, in the limit of large $M$ (heat conduction is neglected) and $Z=1$, overall efficiencies which are at least only one order of magnitude smaller than the Carnot efficiency $\eta_C = \Delta T/T_\infty \approx {\cal O}(0.1-1\:\%)$ (for $\Delta T = 1-10\:\textrm{K}$ and ambient conditions) appear only feasible for channels being a few $\textrm{nm}$ thick. In this context, one has to keep in mind that all the above analysis has been carried out for $H\gg \lambda_{\textrm{D}}$. Alternatively, reducing $\rho$ by confining air bubbles in the liquid domain, i.e. by using a "porous" working fluid such as liquid foams, might be a feasible approach, reducing the heat transported within the bulk of the fluid. Such systems will depend on the addition of surfactants to stabilize the enclosed air pockets. These surfactants may have an effect on the ion distribution so that a more detailed analysis beyond the scope of the present paper is required. Furthermore, the presence of surfactants will generally decrease the value of $\gamma_T$.

In the limit of a small Marangoni number, heat transfer is conduction-dominated and the overall efficiency can be approximated by
\begin{equation}\label{Eq:eff_t2e_conddom} 
\left(\eta^{(\gamma_T)}_{\textrm{t}2\textrm{e},\textrm{max}}\right)_{M \rightarrow 0} = 
\frac{\gamma^2_T \Delta T}{\eta k} \frac{W}{H} \beta_{\|} \frac{Z \sqrt{1-Z}}{(1+\sqrt{1-Z})^2}.
\end{equation}
Thus, in this limit of vanishingly small fluid velocities, it might be beneficial to use -next to a large value of $W/H$- a discontinuous fluid domain since in this case $k$ is small. A similar measure is not effective if, compared to convection, heat conduction in the liquid is negligible since the majority of thermal energy would be transported in each liquid parcel.

\subsubsection{Destillocapillarity-induced streaming}\label{Sec:destillcap}

Surface tension does not depend only on temperature but also on the concentration of surface active components. Hence, the electrokinetic streaming might be induced by a surface stress caused by a gradient in the bulk concentration $c$ of a surfactant. For simplicity, only a single surfactant is considered. In the dilute limit, the activity coefficient 
is approximately equal to unity so that 
\begin{equation}\label{Eq:surf_tens_conc} 
\partial_z \sigma = -\Gamma \partial_z \mu^{(\textrm{s})} \approx - \Gamma R_{\textrm{m}} T \partial_z \textrm{ln}(c),
\end{equation}
where $\mu^{(\textrm{s})}$ is the chemical potential of the surfactant, $R_{\textrm{m}}=8.31\,\mathrm{J/(mol\,K)}$ is the universal gas constant, and $\Gamma$ is the surface excess concentration. If the flow is fully developed and the interface remains flat, the power required to maintain the concentration gradient can be approximated by
\begin{align}\label{Eq:chem_power}
P_{\textrm{ch}} &= a W \Gamma \Big(\overline{w}_{|\textrm{i}} - D^{(\textrm{s})}_{|\textrm{i}} \partial_z \textrm{ln}(\Gamma)\Big) \Delta \mu^{(\textrm{s})} \nonumber \\
&+ W H c \left(\overline{w} - D^{(\textrm{s})} \partial_z \textrm{ln}(c)\right) \Delta \mu^{(\textrm{s})},
\end{align}
where $D^{(\textrm{s})}_{|\textrm{i}}$ and $D^{(\textrm{s})}$ are the surfactant diffusivities at the surface and in the bulk, respectively. A non-soluble surfactant corresponds to $c \rightarrow 0$ so that the conversion efficiency from chemical to mechanical energy is approximately given  by
\begin{equation}\label{Eq:eff_chem2mech} 
\eta_{\textrm{c}2\textrm{m}} = \frac{P_{\textrm{in}}}{P_{\textrm{ch}}} = \frac{1}{1 - \frac{D^{(\textrm{s})}_{|\textrm{i}}}{\overline{w}_{|\textrm{i}}} \partial_z \textrm{ln}(\Gamma)}  .
\end{equation}
In case of maximum efficiency ($\Theta = 1/\sqrt{1-Z}$), one finds
\begin{equation}\label{Eq:eff_diff_conv}
\frac{D^{(\textrm{s})}_{|\textrm{i}}}{\overline{w}_{|\textrm{i}}} \partial_z \textrm{ln}(\Gamma) = \frac{a}{\Gamma} \frac{D^{(\textrm{s})}_{|\textrm{i}} \eta}{R_{\textrm{m}} T W \beta_{\|}} 
\frac{\partial_z \textrm{ln}(\Gamma)}{\partial_z \textrm{ln}(c)} \frac{1}{\sqrt{1-Z}}.
\end{equation}
If surface diffusion is negligible, then $P_{\textrm{ch}} \approx P_{\textrm{in}}$ and the overall conversion efficiency from chemical 
to electric energy is given by (\ref{Eq:eff_m2e_press}). Typically, for the channels with superhydrophobic walls as treated herein, 
this maximum efficiency remains for realistic values of the induced surface charge density of below $10^{-4}\:\textrm{C m}^{-2}$ within 
a few percent \cite{Seshadri:PoF2013}.

\section{Conclusions}\label{sec:conclusions}
In this study, the charge separation and energy conversion in electrokinetic free-surface flow driven by a gradient in surface tension was analyzed. At hand of a simple Couette-type of flow it was shown that there is a qualitative difference between shear- and pressure-driven electrokinetic streaming if charges are only present at non-slipping walls. While the streaming potential generated by Poiseuille-type of flow typically attains a constant value at large channel cross sections, the streaming potential per shear and $\zeta$-potential vanishes at large film heights $H$ according to $(H/\lambda_{\textrm{D}})^{-1}$, where $\lambda_{\textrm{D}}$ denotes the thickness of the Debye layer. At $H \approx {\cal O}(\lambda_{\textrm{D}})$, the decrease of the streaming potential with increasing $H/\lambda_{\textrm{D}}$ is less than for larger $H$. This is caused by the -in this limit more important- contribution stemming from the electro-osmotic counter flow, which is even more pronounced with increasing values of the higher surface charge densities.

For large film heights, the Helmholtz-Smoluchowski limit present in pressure-driven streaming is seen to be (qualitatively) recoverable for shear-driven flow if an electric field is applied perpendicular to the free surface to induce an interfacial charge density. In the limit of small $H \approx {\cal O}(\lambda_{\textrm{D}})$, the streaming potential can be further manipulated within a wide range if the solid walls exhibit a molecular slip of ${\cal O}(\lambda_{\textrm{D}})$. This leads, along with the values and signs of the wall and interfacial $\zeta$-potentials, to a multitude of possible parameter variations. More specifically, it was seen that molecular slip is only beneficial for low to moderate $\zeta$-potentials. This was again traced back to the opposing effect of the electro-osmotically driven counter-flow.

In the limit of an infinitely thin double layer, the findings were compared with a technically more feasible slit channel flow bounded by superhydrophobic walls. To this end, the flow profile and streaming potential were derived when a surface tension gradient along the air-filled micro-structured grooves is used to propel the fluid. It was shown that the final equations governing the electrokinetic streaming are equivalent to those obtained for the Couette-type of flow if a corresponding velocity scale and apparent slip length is used. In this context it was discussed that not only this velocity scale increases with increasing free surface fraction but accordingly also the figure of merit, which in turn implies a higher 
conversion efficiency from mechanical to electric energy.

In the case that the variation of surface tension is temperature-induced, a thermal-to-electric conversion efficiency was derived. For large thermal Marangoni number, the efficiency was seen to be severely limited by the typically small Marangoni coefficient as well as by the large heat capacity of typical electrolytes. It was argued that even employing discontinuous fluid domains may not be helpful to remedy this fundamental problem. By contrast, for small Marangoni number, it might be beneficial to split the fluidic domain into parcels or foam lamellas to reduce the effective thermal and electric conductivity. However, a more detailed study has to be undertaken to quantify the effects of surfactants added for the stabilization of the air pockets. For continuous domains and any Marangoni number, the maximal achievable thermal efficiency is found to be at least an order of magnitude smaller than the Carnot factor, even if the film/channel cross section is not larger than ${\cal O}(\lambda_{\textrm{D}})$. Finally, it was demonstrated that for a concentration-dependent surface tension, conversion efficiencies can be achieved which qualitatively agree with those obtained for pressure-driven charge separation, being typically of ${\cal O} (1\%)$.

The findings are useful for the design and feasibility tests of devices which employ thermocapillarity or destillocapillarity as a means to generate electric voltage from electrokinetic streaming.

\section*{Acknowledgments}

This work was in part supported by the German Research Foundation (DFG) through Cluster of Excellence 259, 'Center of Smart
Interfaces'. Steffen Hardt is acknowledged for fruitful discussion.

\bibliographystyle{elsarticle-num}
\bibliography{references}

\appendix
\section{Solution details of the Laplace equation}\label{sec:appendix_checkAnalytic}

The velocity field between parallel plates containing periodic patches of no-slip and constant-shear regions was estimated using the result by Philip, equation (\ref{Eq:non_dim_vel_field}), for such a flow in an infinite half-plane, $y\geq 0$. No analytical result is known for a finite plate separation. However, the range of validity of this approximation can be assessed numerically. For this, the Laplace equation, $\nabla^2 w =0$, was discretized using the finite element method as implemented in the commercial code Comsol Multiphysics. By symmetry, the computational domain can be restricted to a unit cell indicated by the gray area in figure \ref{Fig:schematic_microstructure}. The boundary conditions are $w(x,0) = 0$ for $0<x<(1-a)W$ (at the solid wall) and $\eta \partial_y w(x,0) = -\tau$ for $(1-a)W<x<W$ (at the constant-shear surface). On all other boundaries symmetry conditions apply, i.e. $(\vec{n} \cdot \vec{\nabla}) w = 0$ with $\vec{n}$ being the outward normal at the boundary. The average velocity, $(HW)^{-1}\int_0^H\!\!\int_0^W\!\! w\,dx\,dy$, normalized with the analytic value corresponding to Philip's solution, $\beta_\| W \tau/\eta$, is tabulated in table \ref{Tab:NumNormalisedVelocity} for different values of the free surface fraction, $a=B/W$, and aspect ratio, $H/W$. It is evident from the table that for $H/W=1$ the numerically obtained results deviate by only ${\cal O}(10^{-3})$ from the corresponding analytical result and even for $H/W=0.75$ the agreement is ${\cal O}(10^{-2})$.

\begin{table}
\center
\begin{tabular}{|c||c c c c|}
	\hline
	$a=$	& \multicolumn{4}{c|}{$H/W$}\\
	\cline{2-5} 
	$B/W$		& 0.5 & 0.75 & 1 & 1.5 \\
	\hline\hline
	0.25	& 1.024 & 1.005 & 1.001 & 1.000 \\
	0.5	& 1.064 & 1.013 & 1.003 & 1.000 \\
	0.75	& 1.069 & 1.014 & 1.003 & 1.000 \\
	0.8	& 1.063 & 1.013 & 1.003 & 1.000 \\
	0.9	& 1.047 & 1.009 & 1.002 & 1.000 \\
	0.95	& 1.036 & 1.007 & 1.001 & 1.000 \\\hline
\end{tabular} 
\caption{\label{Tab:NumNormalisedVelocity} Numerical values for the normalized average velocity, $(\beta_\| HW^2 \tau/\eta)^{-1}\int_0^H\!\!\int_0^W\!\! w\,dx\,dy$, for different values of the free surface fraction, $a=B/W$, and aspect ratio, $H/W$, as obtained using a finite element discretization.}
\clearpage
\end{table}

\begin{figure}
\center
\includegraphics[scale=0.95]{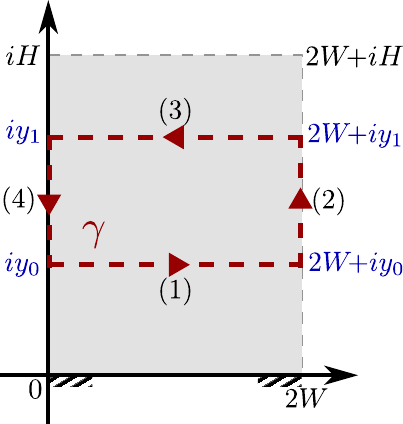}
\caption{\label{Fig:IntegrationContour} Sketch of the integration contour, $\gamma$. Due to periodicity, the integration along sections $(2)$ and $(4)$ cancel.}
\end{figure}

For the analysis presented in the main text, flow rates and line averages of the velocity field are needed. Normalized with the length or area of the integration region, these averages turn out to be identical, attesting the relevance of table \ref{Tab:NumNormalisedVelocity}. In fact, even for finite $H$, one can show that
\begin{equation}\label{Eq:lineIntegrals}
W^{-1}\int_0^W\! w(x,y_0)dx=W^{-1}\int_0^W\! w(x,y_1)dx
\end{equation}
for any $0\leq y_0 < y_1 \leq H$; thus line averages and the flow rate are inherently linked. A sketch of a proof of this relation is as follows: since $w$ is harmonic, $\nabla^2 w=0$, it is the imaginary part of a holomorphic function $f(x+iy)=v(x,y)+iw(x,y)$ with $v,w:\mathbb{R}^2\to\mathbb{R}$, \cite{Lawrentjew:1967}. By Cauchy's integral theorem, $\Im[\oint_\gamma f({\xi})d\xi]=0$ for any closed path $\gamma$. Choose $\gamma$ as the rectangle with vertices $\xi=iy_0$, $2W+iy_0$, $2W+iy_1$, $iy_1$ as shown in figure \ref{Fig:IntegrationContour}. Due to the Cauchy-Riemann conditions, $v(x,H)$ is constant since $\partial_x v|_{y=H} =-\partial_y w|_{y=H}=0$; similarly, $v(0,y)$ and $v(2W,y)$ are constant since $\partial_x w|_{x=0}=0=\partial_x w|_{x=2W}$. Thus not only $w(x,y)=w(x+2W,y)$ is periodic in $x$, but so is $v$ and thus $f$. The line integrals on the legs with constant $x$ thus cancel, which completes the proof.

\end{document}